\documentclass[graybox, envcountchap]{svmult}

\usepackage{mathptmx}        
\usepackage{helvet}          
\usepackage{courier}         

\usepackage{makeidx}         
\usepackage{graphicx}        
\usepackage{multicol}        
\usepackage[bottom]{footmisc}

\makeindex             

\usepackage{amsmath}
\usepackage{amssymb}
\usepackage{lscape}
\usepackage{multirow}

\def\gapp{\ifmmode\stackrel{>}{_{\sim}}\else$\stackrel{<}{_{\sim}}$\fi}
\def\gsim{\lower.5ex\hbox{\gtsima}}
\def\gtsima{$\; \buildrel > \over \sim \;$}
\def\lapp{\ifmmode\stackrel{<}{_{\sim}}\else$\stackrel{<}{_{\sim}}$\fi}
\def\lsim{\lower.5ex\hbox{\ltsima}}
\def\ltsima{$\; \buildrel < \over \sim \;$}

\newcommand\apgt{\ {\raise-.5ex\hbox{$\buildrel>\over\sim$}}\ }
\newcommand\aplt{\ {\raise-.5ex\hbox{$\buildrel<\over\sim$}}\ }
\begin{document}
\pagestyle{empty}
\frontmatter

\include{dedic}
\include{foreword}
\include{preface}

\mainmatter

%
%
%
\setcounter{chapter}{12}
\titlerunning{BSSs in globular clusters: observations, statistics and physics}
\title{Blue Stragglers in Globular Clusters: Observations, Statistics and Physics}


\author{Christian Knigge}
\institute{Christian Knigge \at Physics \& Astronomy\\
University of Southampton\\
Southampton SO17 1BJ, UK, \email{C.Knigge@soton.ac.uk}}

\maketitle
\label{Chapter:Knigge} 

\abstract*{This chapter explores how we might use the
  observed {\em statistics} of blue stragglers in globular clusters to
  shed light on their formation. This means we will touch on topics
  also discussed elsewhere in this book, such as the discovery and
  implications of bimodal radial distributions and the ``double
  sequences'' of blue stragglers that have recently been found in 
  some clusters. However, we will focus 
  particularly on the search for a ``smoking gun'' correlation between
  the number of blue stragglers in a given globular cluster and a physical cluster 
  parameter that would point towards a particular formation
  channel. As we shall see, there is little evidence for
  an intrinsic correlation between blue straggler numbers and stellar
  collision rates, even in dense cluster cores. On the other hand,
  there {\em is} a clear correlation between blue straggler numbers
  and the total (core) mass of the cluster. This would seem to point
  towards a formation channel involving binaries, rather than
  dynamical encounters. However, the correlation between blue
  straggler numbers and actual binary numbers -- which relies on
  recently determined empirical binary fractions -- is
  actually {\em weaker} than that with core mass. We explain how this
  surprising result may be reconciled with a binary formation
  channel if binary fractions depend almost uniquely on core
  mass. If this is actually the case, it would have significant
  implications for globular cluster dynamics more generally.}

\section{Straw-Man Models for Blue Straggler Formation}
\label{knisec:straw}

In order to gain some intuition, let us start by considering the two
simplest distinct formation channels\index{formation channel} for blue stragglers in globular
clusters\index{globular cluster}. First, blue stragglers may form in the same way in clusters
as they do in the Galactic field, i.e. via mass transfer\index{mass transfer} or
coalescence\index{coalescence} in binary systems\index{binary system}. In this case, we
may expect the number of blue stragglers in any given cluster
($N_{BSS}$) to scale with the number of binary stars in the
cluster ($N_{bin}$), 
\begin{equation}
N_{BSS} \propto N_{bin} \propto f_{bin} M_{tot},
\end{equation}
where $f_{bin}$ is the fraction of binaries among the cluster members 
and $M_{tot}$ is the total mass of the cluster. In reality, $f_{bin}$
should really be the fraction of {\em close} binaries\index{close binary} (since only
these can be the progenitors\index{progenitor} of blue stragglers), but let us
assume for the moment that these two quantities track each other, so
that we can ignore this subtlety.

The second possibility is that blue stragglers in
globular clusters form primarily via dynamical encounters\index{dynamical encounter}. Here, the
simplest possibility is that the most important encounters are direct
collisions\index{collision} between two single stars. In this case, the number of blue
stragglers should scale with the {\em 1+1 collision rate}
($\Gamma_{coll,1+1}$), which is determined by the conditions in the
dense cluster core via
\begin{equation}
N_{BSS} \propto \Gamma_{coll,1+1} \propto R_c^3 n_c^2 \sigma_c^{-1}.
\end{equation}
Here, $R_c$ is the core radius of the cluster, $n_c$ is the stellar
density\index{stellar density} in the core, and $\sigma_{c}$ is the core velocity dispersion\index{velocity dispersion}
(which is a measure of the characteristic speed at which stars move in
the core\index{core}). 

\begin{figure}
\begin{center} 
\includegraphics[width=89mm]{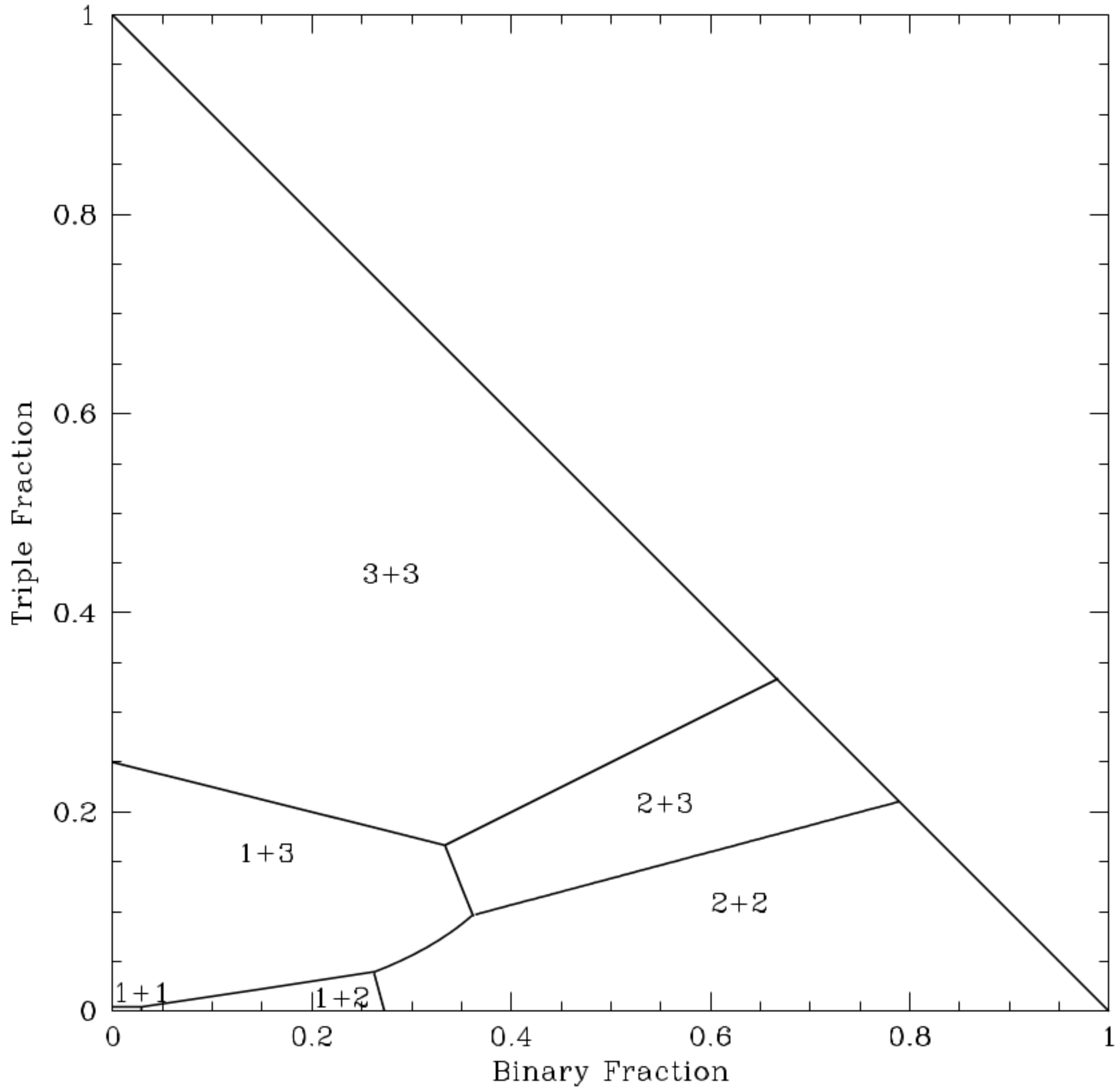}
   \caption{The ``phase diagram'' of dynamical encounters. For any
     stellar population described by a particular combination of
     binary fraction and triple fraction, it is possible to determine
     which type of dynamical encounter will dominate. The plot above
     shows the regions of parameter space dominated by the various
     different encounters, assuming a particular set of characteristic
     binary and triple parameters. Reproduced from Figure~1 of Leigh
     \& Sills \cite{LS11}, {\em An analytic technique for constraining the
     dynamical origins of multiple star systems containing merger
     products}, MNRAS, 410, 2370.}
   \label{triple_phase}
   \end{center}
\end{figure}

These scaling relations are clearly vast
oversimplifications. Perhaps most obviously, {\em both} channels may
produce significant number of blue stragglers in globular
clusters. However, more importantly, even the very distinction between
binary and dynamical formation channels is something of a false
dichotomy. After all, the close binaries that are the progenitors of
blue stragglers in the binary evolution channel may themselves have
been formed or hardened by previous dynamical encounters (e.g. \cite{Hut92}). Similarly, it is not at all obvious that the total rate of
stellar collisions should be dominated by encounters between single
stars. In fact, Leigh \& Sills \cite{LS11} show that the rate of dynamical {\em 
encounters} is dominated by binaries (or even triples) even in
environments with only modest binary (or triple) fractions
(Figure~\ref{triple_phase}). The probability of an actual stellar {\em
collision} occurring in such encounters is discussed by \cite{LG12}. Thus, in reality, the binary channel may involve
dynamical encounters, and the dynamical channel may involve binaries.

Does this mean that our simple straw-man models are useless? Not
at all. First, the most extreme cases one can envision within the two
channels {\em are} basically distinct. If blue straggler formation
is dominated by mass transfer in or coalescence of {\em primordial}
binaries\index{primordial binary} that have not been affected (much) by dynamical encounters,
$N_{BSS}$ will scale with $N_{bin}$ and {\em not} with
$\Gamma_{coll,1+1}$. Conversely, if the dominant channel are really
single-single encounters, then $N_{BSS}$ will scale with
$\Gamma_{coll,1+1}$ and {\em not} with $N_{bin}$. Second, and perhaps
more importantly, we might expect the basic scaling relations to be
valuable even if binaries are affected by encounters and collisions
involve binaries. This is particularly easy to see for the binary
channel. Here, the relationship $N_{BSS} \propto N_{bin}$ should
presumably hold regardless of how the relevant binaries were
formed (so long as our definition of $N_{bin}$ does, in fact, refer to
the ``relevant'' binary population, which may be a significant challenge in
practice). For example, suppose that most of the close binaries that
evolve into blue stragglers have been previously hardened in three-body
encounters\index{three-body
encounter}. In this case, we would expect $N_{BSS} \propto N_{bin}
\propto \Gamma_{1+2}$, where $\Gamma_{1+2}$ is the 1+2 {\em encounter}
rate. Similarly, if blue stragglers are predominantly formed by direct
collisions occurring during 1+2 encounters, we would expect
$N_{BSS} \propto P_{coll,1+2} \propto P_{coll,1+2} \Gamma_{1+2}
\propto f_{bin} (a_{bin}/R_*) P_{coll,1+2} \Gamma_{coll,1+1}$, 
where $P_{coll,1+2}$ is the probability of a physical collision occurring 
during a 1+2 encounter, $a_{bin}$ is the characteristic binary separation and $R_*$ is the characteristic stellar radius; see \cite{Leigh13} for expressions linking the various encounter rates.

These specific examples show that, at the most basic level, the
straw-man relations should remain roughly valid, even if reality is
more complex than the limiting cases they formally represent. If blue
straggler formation mostly involves binaries, we expect a
scaling with $N_{bin}$; if it mostly involves encounters, we expect a
scaling with $\Gamma_{coll,1+1}$. If binaries and dynamics work in
tandem, $N_{bin}$ and $\Gamma_{coll,1+1}$ will simply be less distinct
quantities.

Two final, technical points are worth noting here. First, $N_{bin}$
and $\Gamma_{coll,1+1}$ will {\em always} be statistically correlated,
even if binaries are primordial and encounters dominated by single
stars. After all, there are both more encounters and more binaries in
an environment containing more stars. Davies, Piotto \& De Angeli \cite{DPd04}  show that this effect induces a scaling of $\Gamma_{coll,1+1}
\propto M_{tot}^{3/2}$, and $N_{bin} \propto f_{bin} M_{tot}$. This
needs to be kept in mind when comparing $N_{BSS}$ to either of these
quantities. 

Second, two different conventions are sometimes used in statistical
studies of blue stragglers. The first and simplest is to use raw {\em
numbers}, $N_{BSS}$, corrected (if necessary) for partial coverage of the
relevant cluster. The second is to use blue straggler {\em
frequencies}, $N_{BSS}/N_{ref}$, where $N_{ref}$ refers to the
number of some reference population (e.g. horizontal branch stars) in
the same field of view. It is important to understand that this 
difference matters. In particular, the straw-man scalings we have derived
above hold only for blue straggler numbers. In
the collision scenario, blue straggler frequencies\index{blue straggler frequency} should scale
with the {\em specific} encounter rate, $\Gamma_{1+1}/M_{tot}$. In the
binary scenario, blue straggler frequencies should scale simply with
the binary fraction, $f_{bin}$.

\section{All Theory is Grey:
Binary Coalescence and Dynamical Encounters in Practice}
\label{knisec:grey}

It is interesting to ask at this point whether there is any
{\em empirical} evidence that the physical mechanisms we are invoking
in our two basic blue straggler formation channels actually occur in
nature. Let us first consider the binary
channel. {\em Mass transfer}\index{mass transfer} is, of course, a well-established process
in many close binary systems, including X-ray binaries\index{X-ray binary} (in which the
accretor is a neutron star\index{neutron star} or black hole\index{black hole}), cataclysmic variables\index{cataclysmic variable} (in
which the accretor is a white dwarf\index{white dwarf}) and Algols\index{Algol system} (in which the accretor
is a main sequence star). But is there also evidence that full
coalescence\index{coalescence} can occur?

As it turns out, there is. It has been known for quite a long time
that some binary system, and in particular the eclipsing W UMa stars\index{W UMa star},
are {\em contact binaries}\index{contact binary}, in which {\em both} binary components
overfill their respective Roche lobes\index{Roche lobe}. In many such systems, the
predicted time scale for full coalescence is much shorter than a
Hubble time. W~UMa binaries are therefore obvious progenitor
candidates for apparently single blue stragglers in the Galactic
field. In fact, quite a few blue stragglers in globular clusters are
known to be (in) W UMa binaries.

\begin{figure}
\begin{center} 
\includegraphics[width=58mm]{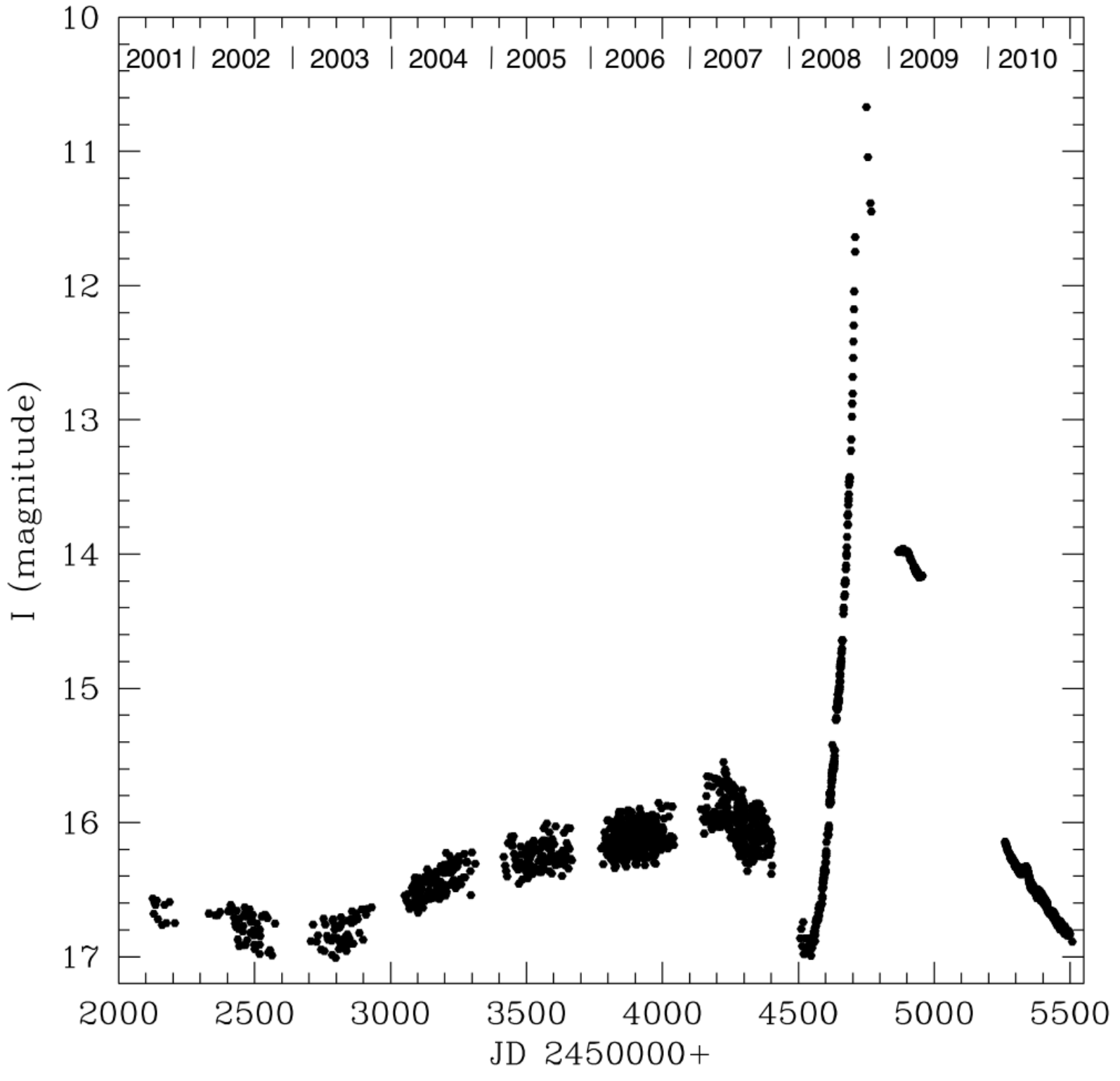}
\includegraphics[width=58mm]{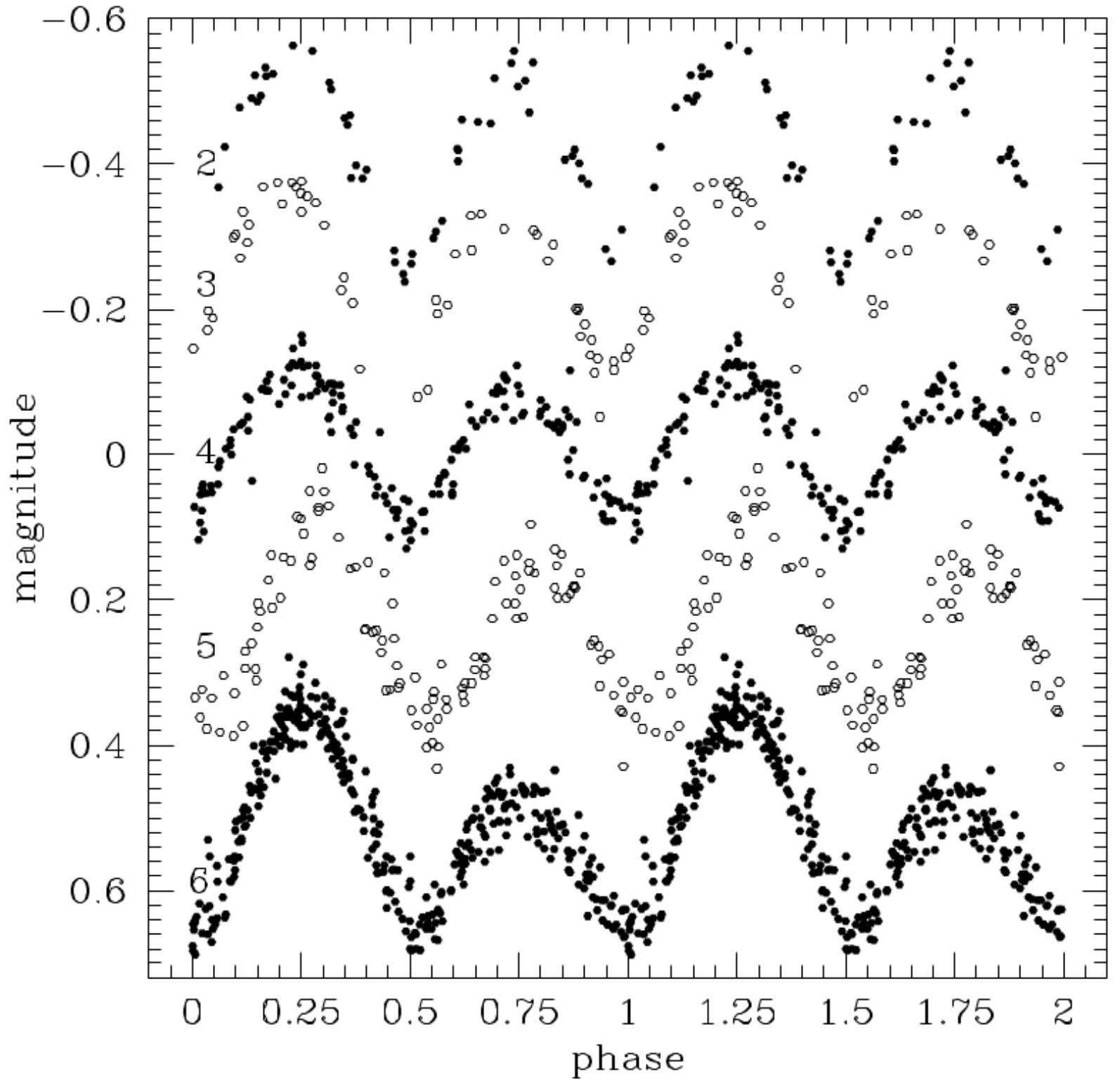}
   \caption{Left: The long-term OGLE light curve of ``Nova'' Sco 2008 = V1309
     Sco. Right: The phase-folded pre-eruption light curves of V1309
     Sco for the 2002-2006 OGLE observing seasons. Reproduced from
     Figure~1 and 3 of Tylenda et al. \cite{Ty11}, {\em V1309 Scorpii: Merger of a
     Contact Binary}, A\&A, 528, A114.}
   \label{tylenda_light}
   \end{center}
\end{figure}

\begin{figure}
\begin{center} 
\includegraphics[width=89mm]{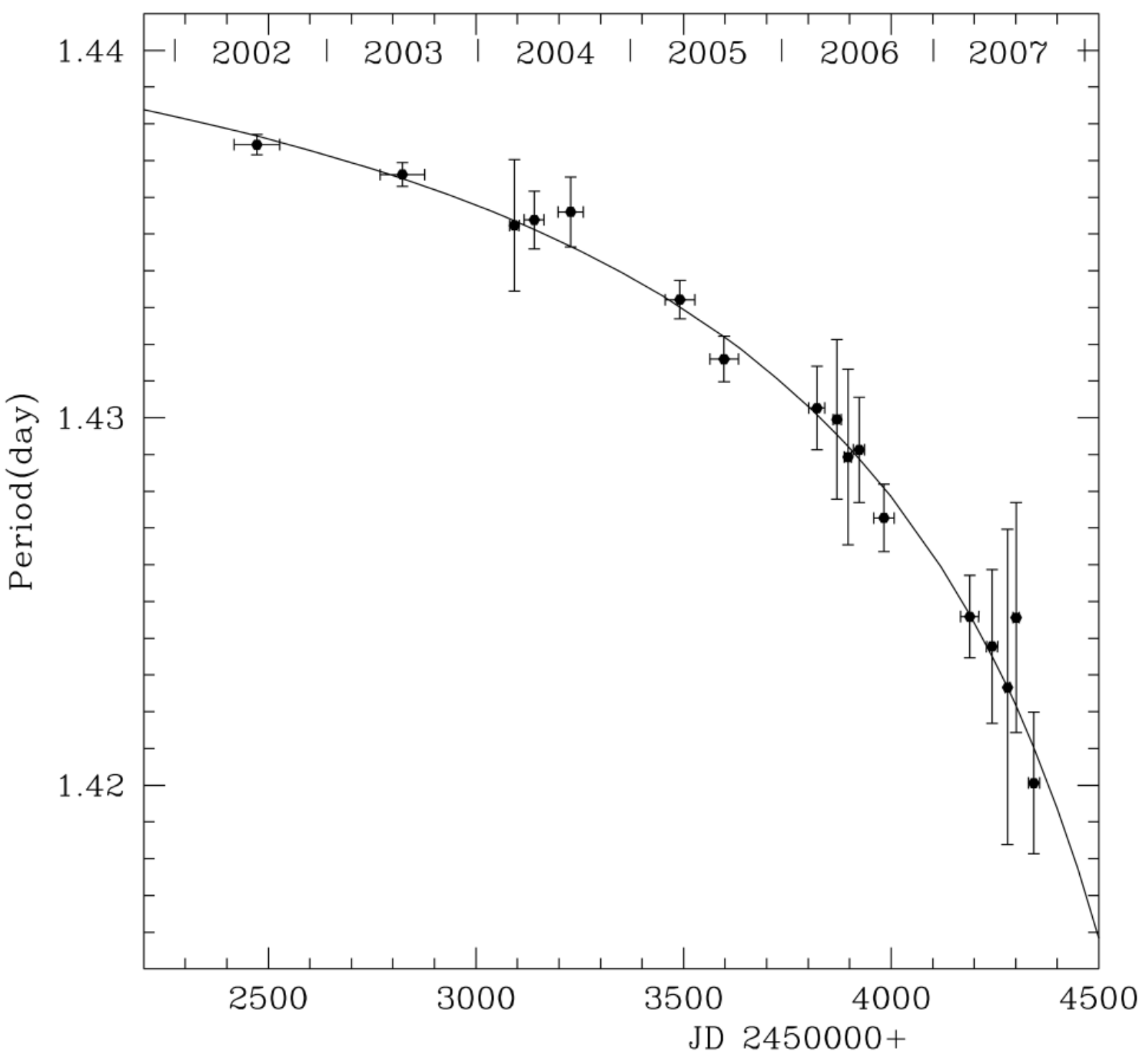}
   \caption{The change in the orbital period of V1309 Sco in the years
     leading up to the 2008 eruption.  Reproduced from
     Figure~2 of Tylenda et al. \cite{Ty11}, {\em V1309 Scorpii: Merger of a
       Contact Binary}, A\&A, 528, A114.}
   \label{tylenda_pdot}
   \end{center}
\end{figure}

However, we can actually do even better. In 2008, astronomers in Japan
and China discovered an apparent nova in the constellation Scorpius
\cite{Nakano2008}. Follow-up observations \cite{Mason10} quickly
revealed that Nova Sco 2008 (aka V1309 Sco\index{V1309 Sco}) was quite an unusual
transient and probably related to the rare class of ``red novae''\index{red nova}
(like V838 Mon\index{V838 Mon}). By a huge stroke of luck, the nova\index{nova} happened to lie in the 
footprint of the OGLE microlensing survey\index{OGLE microlensing survey} \cite{Udalski2003}. The pre-
(and post-)eruption OGLE data of V1309 Sco is astonishing \cite{Ty11}. Not only does it provide an exquisite light curve of what
turns out to be a $\simeq$6 mag eruption, but it also reveals that
{\em the system was a $P_{orb} \simeq 1.4$~d W UMa contact binary
prior to the outburst} (Figure~\ref{tylenda_light})! In fact, the OGLE data is
good enough to provide several accurate measurements of the orbital
period in the lead-up to the eruption (Figure~\ref{tylenda_pdot}). These
measurements show that $P_{orb}$ decreased significantly in just the
$\simeq$6 years leading up to the outburst. The implication is that
{\em Nova Sco 2008 represents a binary coalescence event caught in
real time!}


What about dynamical encounters\index{dynamical encounter} or direct stellar collisions\index{stellar collision} in
globular clusters? No such event has been unambiguously observed in
real time to date. This is not surprising given the low frequency and
short duration of such events. There is nevertheless very
strong empirical evidence that some exotic stellar populations in
globular clusters are preferentially formed in dynamical encounters. 

\begin{figure}
\begin{center} 
\includegraphics[width=119mm]{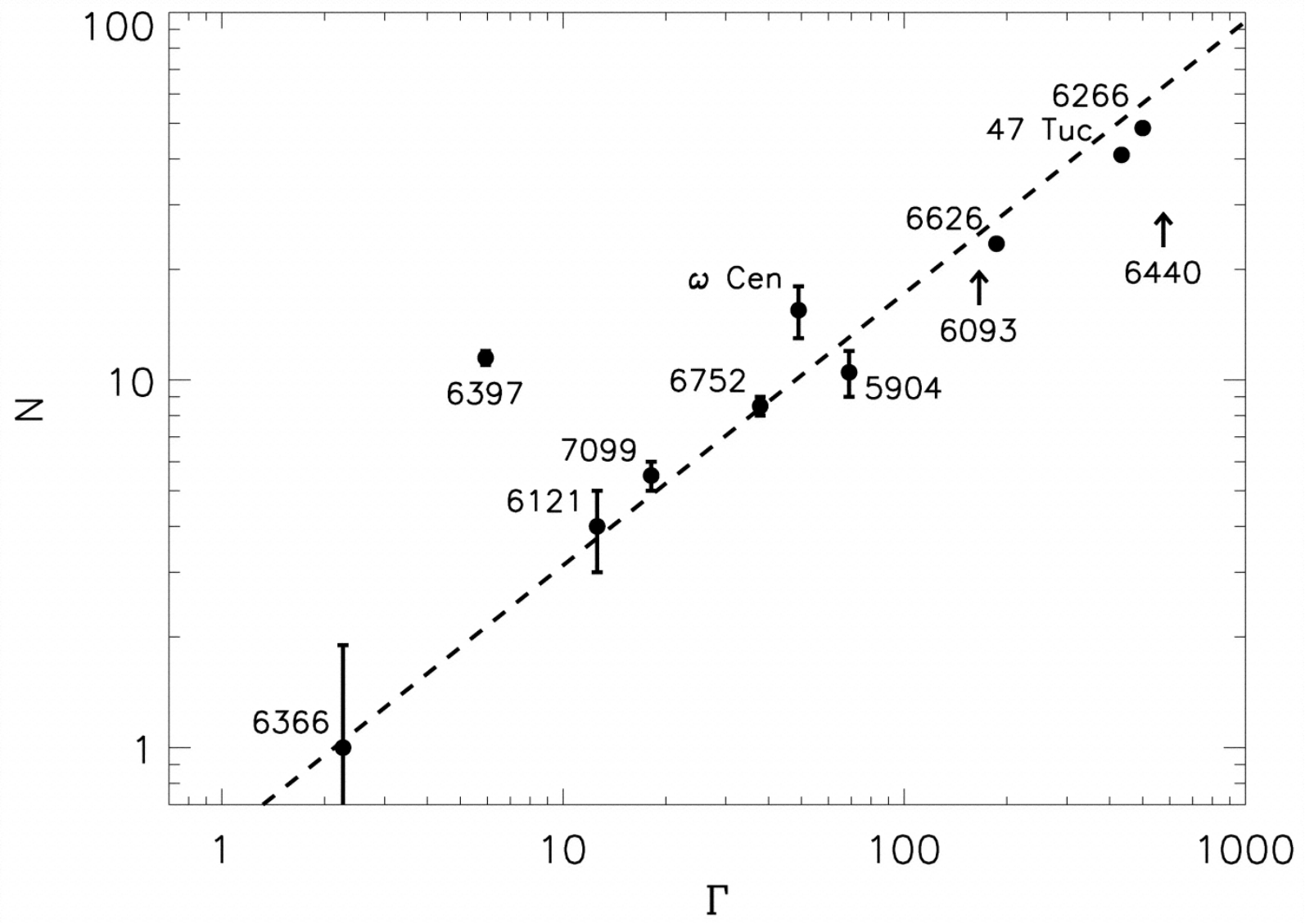}
   \caption{The number of X-ray sources detected in globular cluster
     above $L_x \simeq 4 \times 10^{30}~{\rm erg s^{-1}}$ versus the
     normalised collision rate of the cluster. Reproduced by permission 
     of the AAS from Fig 2
     of Pooley et al. \cite{Po03}, {\em Dynamical Formation of Close Binaries
       in Globular Clusters}, ApJL, 591, L141.}
   \label{pooley}
\end{center}
\end{figure}

It has been known since the 1970s that bright low-mass
X-ray binaries\index{low-mass
X-ray binary} are overabundant by a factor of $\simeq 100$ in globular
clusters, relative to the Galactic field. This was quickly ascribed to
the availability of unique {\em dynamical} formation channels in
globular clusters, such as 1+1 tidal captures\index{tidal capture} \cite{Katz75,Clark75,FPR75}. However, observational confirmation of
this idea required much more powerful X-ray telescopes\index{X-ray telescope} and took nearly 
another three decades. The first convincing empirical case was made by
Pooley et al. \cite{Po03}, who showed that the number of moderately bright
X-ray sources\index{X-ray source} in a given cluster (which are dominated by a variety of
accreting compact binaries\index{compact binary}) exhibits a clear scaling with the
predicted dynamical collision rate\index{dynamical collision rate} in the cluster
(Figure~\ref{pooley}). 

Pooley \& Hut \cite{PH06} later showed that, in high-collision-rate
clusters, this scaling holds independently for both 
low-mass X-ray binaries and cataclysmic variables. However, there is
also tentative evidence that, in low-collision-rate clusters, the
number of these sources may instead scale with cluster mass. This
suggests that the evolution of (primordial?) binaries may be
sufficient to produce the few X-ray sources observed in such
clusters. Our simple straw-man models thus do a rather good job in
accounting for the observed number of X-ray binaries across the full
range of Galactic globular clusters.

\pagebreak

\section{The Search for the Smoking Gun Correlation I:\\
The Near Constancy of Blue Straggler Numbers}
\label{knisec:smoke1}
\index{correlation}

Let us return to blue stragglers. The first reasonably complete
catalogue of blue stragglers in Galactic globular clusters was
constructed and analysed by Piotto et al. \cite{Pio04}. This catalogue was
based on an HST/WFPC2 survey\index{HST/ACS survey} that provided V and I colour-magnitude
diagrams for 74 clusters \cite{Pio02}. Blue stragglers could
be reliably selected in 56 of these clusters, yielding a total sample
of nearly 3000 stars.

The results obtained by Piotto et al. \cite{Pio04} were surprising. Most
importantly, they found {\em no} correlation between the 
frequency of blue stragglers and the cluster collision
rate. Moreover, they found a weak {\em anti-correlation} between blue
straggler frequency and cluster luminosity\index{cluster luminosity} (i.e. total mass\index{cluster mass}). One
potentially confounding issue in their analysis is that, it is
somewhat unnatural to correlate blue straggler frequencies against
collision rate\index{collision rate} and total mass. As emphasised in Section~\ref{knisec:straw}, it
is the {\em number} of blue stragglers that should scale linearly with
these quantities in our straw-man models, not their frequency.

\begin{figure}
\begin{center} 
\includegraphics[width=119mm]{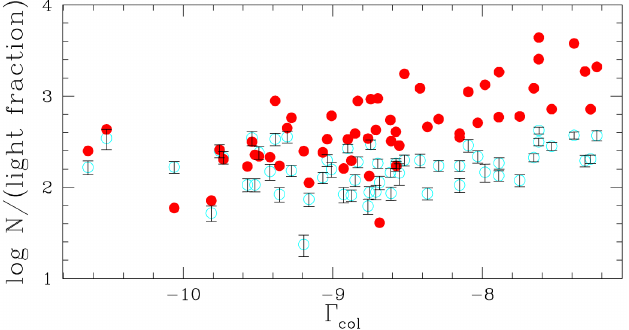}
\includegraphics[width=119mm]{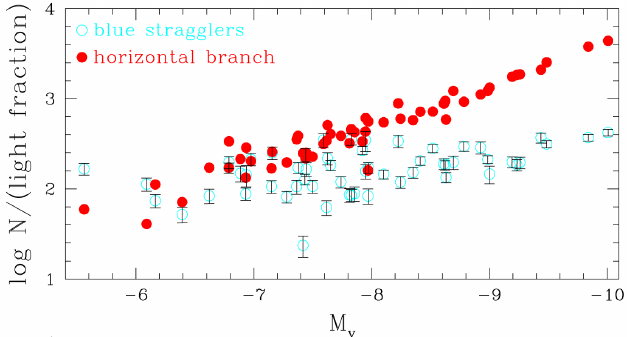}
   \caption{{\em Top panel:}The observed number of blue stragglers and horizontal
     branch stars as a function of stellar collision rate. {\em Bottom
       panel:} The estimated number of blue stragglers and horizontal
     branch stars as a function of the absolute magnitude of the
     cluster. Reproduced from Fig 1 of Davies et al. 
     \cite{DPd04}, {\em Blue Straggler Production in Globular Clusters},
     MNRAS, 348, 129.}
   \label{davies}
   \end{center}
\end{figure}

As it turns out, however, this issue is not the main cause of the
unexpected results. Indeed, the same data base was re-analysed and
interpreted by Davies et al. \cite{DPd04}, who showed that
blue straggler {\em numbers} also do not correlate significantly with
collision rate (Figure~\ref{davies}; top panel). They also argued that blue
straggler numbers are largely independent of total cluster
mass/luminosity, although an inspection of their figure suggests that there
may be a mild, positive, but sub-linear correlation between these
quantities (Figure~\ref{davies}; bottom panel).

The number of horizontal branch stars\index{horizontal branch star} does scale linearly with cluster
mass or luminosity, as one would expect for a ``normal'' stellar
population (Figure~\ref{davies}; top panel). More interestingly, the number of
horizontal branch stars actually also correlates with the collision
rate (Figure~\ref{davies}; bottom panel). This correlation is no doubt
induced by the 
intrinsic correlation between cluster mass and collision rate
(Section~\ref{knisec:straw}). Indeed, the scaling between horizontal branch
numbers and collision rate is broadly in line with the relation we
would expect in this case, i.e. $N_{HB} \propto
\Gamma_{coll,1+1}^{2/3}$. But this only highlights the central mystery:
apparently blue stragglers exhibit a weaker correlation with collision
rate than horizontal branch stars, a population that is certainly not
produced in dynamical encounters. 

\section{Do Clusters Deplete their Reservoir of Binary Blue Straggler Progenitors?}
\label{knisec:burn}

Davies et al. \cite{DPd04} suggested an interesting
interpretation for the near constancy of blue straggler numbers
(Figure~\ref{davies}). Their idea invokes a combination of binaries
and dynamical encounters\index{dynamical encounter}. Specifically, they consider  
a binary\index{binary star} mass transfer scenario\index{mass transfer scenario} in which blue stragglers
are formed when the primary star leaves the main sequence, expands and
fills its Roche lobe\index{Roche lobe}. This initiates mass transfer onto the secondary,
which can then be converted into a blue straggler. Davies et al. note
that, in dense globular clusters, each binary is likely to undergo
many encounters with single stars. In each such encounter, the most
likely outcome is the ejection of the least massive star, so these
encounters strongly affect the mass distribution of the binary
population. This, in turn, affects the ability of this population to
form blue stragglers, since only systems with primaries close to the
turn-off mass are viable progenitors. 

The central argument put forward by Davies et al. \cite{DPd04}  is
that, in high collision rate clusters\footnote{Strictly speaking, we are talking here about clusters with high
{\em specific} collision rates, i.e clusters in which {\em each binary}
undergoes many encounters.}, relatively massive stars near the cluster turn-off mass are likely
to have exchanged into binaries well before the present day. Such
clusters may therefore have used up their blue straggler
binaries by the present day and may thus now be deficient in blue
stragglers derived from the binary channel. On the other hand, blue
stragglers formed via direct stellar collisions should be more
numerous in these clusters. Davies et al.  therefore suggest
that these two effects broadly cancel. This would imply that
binary-derived blue stragglers dominate in low-collision-rate
clusters, while collisional blue stragglers dominate in
high-collision-rate clusters, even though the absolute numbers in both
types of clusters are more or less the same. 

In support of this argument, Davies et al.  carried out a simple
simulation. In this, a set of initial (``primordial'') binaries\index{primordial binary} was
created, in which the mass of each binary component is drawn
independently from a simple initial mass function \cite{Egg89,Egg90}. Each binary was then subjected to a series of exchange
encounters with single stars whose masses are also drawn from the same
IMF. In each encounter, the least massive of the three stars involved
was ejected, and the remaining two assumed to remain as a binary
system. After each encounter, a system was labelled as a blue
straggler if the mass of its primary satisfied 
$0.8~{\rm M_{\odot}} < M_1 < 0.816~{\rm M_{\odot}}$. This corresponds to
the range of turn-off masses\index{turn-off mass} over the last 1 Gyr -- a typical blue
straggler life time -- for a typical Galactic globular cluster. 

\begin{figure}
\sidecaption
\includegraphics[width=69mm]{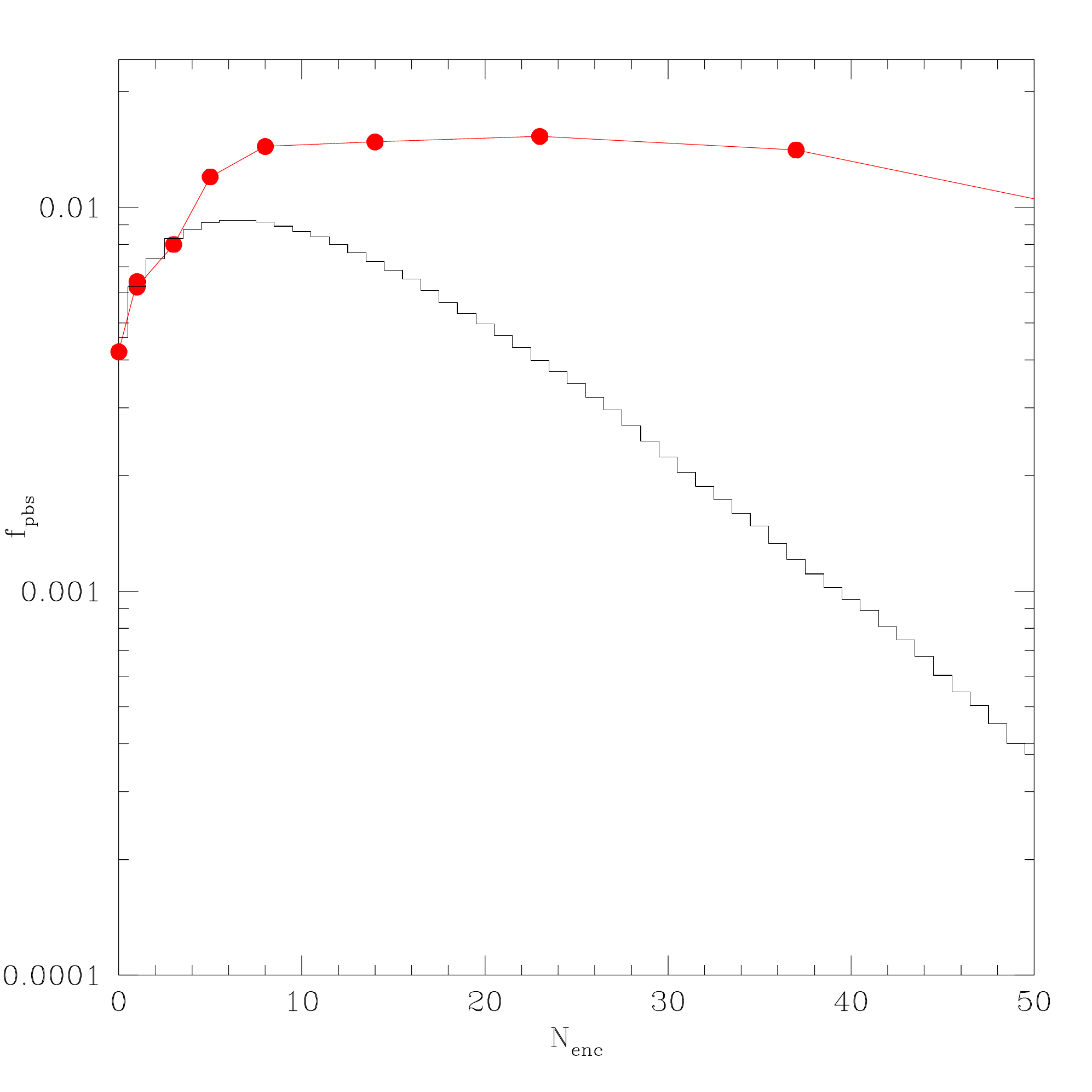}
   \caption{The fraction of ``blue stragglers'' among a
     simulated binary population versus the number of encounters with
     single stars the binary has undergone. Blue stragglers are
     defined as binaries with primaries near the present-day turn-off
     mass. The black line corresponds to the case where the masses of
     the initial binary components, as well as the masses of the
     single stars the binary encounters, are drawn from an
     unrestricted initial mass function. This is the case considered
     by Davies et al. \cite{DPd04} . The red line corresponds to the case
     where, at each encounter, all stellar masses are replaced with a
     suitable compact object mass if they exceed the appropriate
     turn-off mass at the time of the encounter (see text for
     details). Note that while the black line declines fairly quickly towards
     large $N_{enc}$, the red line does not.}
   \label{davies_check}
\end{figure}

In order to gain some insight, we have repeated their simulation. The
black histogram in Figure~\ref{davies_check} shows how
the fraction of blue stragglers among the simulated binaries,
$f_{pbs}$, depends on the number of encounters a binary has undergone,
$N_{enc}$. In agreement with their results, we see that $f_{pbs}$
initially increases as the number of 
encounters goes up, but then peaks at $N_{enc} = 6$ and declines
monotonically towards larger $N_{enc}$. The expected number of
encounters in the highest collision rate clusters is larger than six
over the cluster lifetime, so these results appear to
suggest that binary-derived blue stragglers will indeed be rare in
such clusters today.

There are, however, problems with this scenario. For example, one
would still expect a scaling of blue straggler numbers with
$\Gamma_{coll,1+1}$ at least for high-collision-rate clusters (which
is not really observed). Moreover, since exchange encounters are
required to produce even binary-derived blue stragglers in this
model, we might actually still expect a scaling with
$\Gamma_{coll,1+1}$ for these objects (via an intrinsic scaling with
$\Gamma_{1+2}$; see Section~\ref{knisec:straw}). 

However, the most fundamental problem with the simulation is that it 
assumes that all stellar masses are available for all
encounters. In reality, each successive encounter represents a later time
in the evolution of the cluster, so stars above the main sequence
turn-off corresponding to this time will already have evolved off the
main sequence. In order to test if this effect matters, we have
repeated the simulation once more, but this time with a rough model
for stellar evolution. 

In this new calculation,  we first estimate the typical time interval
between encounters for a given set of representative binary and
cluster parameters. This provides the time step for the simulation. 
We then evolve the binary population forward by allowing each binary
to encounter a single star at each time step. At this point, we first
check if the primary has turned off the main sequence since the last
time step. If so, we assume that mass transfer has already started and
that no further exchange encounters will take place. If not, we
once again ask if an exchange encounter will happen. However, we
now also first check if the single star the binary has encountered has
turned off the main sequence. If so, we replace its mass with that of
the relevant compact object\footnote{If $M_i > 18$~M$_{\odot}$, we assume the star has turned
into a black hole, so that $M_f = 10$~M$_{\odot}$; if $7$~M$_{\odot} < M_i
< 18$~M$_{\odot}$, we assume the star has turned in a neutron star, so
that $M_f = 1.4$~M$_{\odot}$; finally, if $M_{to}(t) < M_i < 7$~M$_{\odot}$,
we assume the star has become a white dwarf, so that $M_f = 0.5
$~M$_{\odot}$. These mass ranges, as well as the main sequence lifetimes,
are estimated using SSE \cite{HPT00}.}.
Once the present day is reached, we calculate the fraction of blue
stragglers in the same way as before (but excluding systems with white
dwarf secondaries). We then carry out the same
calculation for a wide range of assumed cluster densities (and hence
collision rates), with each density corresponding to a different 
number of encounters between the birth of the cluster and the present
day. 

The results of this modified simulation are shown by the red line in
Figure~\ref{davies_check}. With our simplistic treatment of stellar evolution
included, $f_{pbs}$ now rises fairly quickly up to $N_{enc} \simeq 10$
and then stays nearly constant out to at least $N_{enc} \simeq 60$. 
The absence of a sharp decline towards high $N_{enc}$ is actually easy
to understand. In the original simulation, where all 
stellar masses are available in all encounters, the overall binary
population quickly becomes dominated by systems with main sequence
primaries more massive than 0.816~M$_{\odot}$. It is the increasing
dominance of these systems that fundamentally causes the decline in
$f_{pbs}$ towards larger $N_{enc}$. But this is of course unphysical,
since such systems should not exist at the present day. In the revised
simulation, $f_{pbs}$ stays high because the most massive stars in the
cluster (ignoring the extremely rare neutron stars and black holes)
are now always stars with masses near the turn-off mass. 

Davies et al. \cite{DPd04}  emphasised that their simulation
ignored several important physical effects, and this is still true of
our revised simulation as well. Some of these effects are discussed in
more detail in Chapter 9. Nevertheless,
the results in Figure~\ref{davies_check} suggest that the depletion of
blue straggler binaries in high collision rate clusters may not offer
as natural an explanation for the observed blue straggler numbers as
previously envisaged.

\section{The Search for the Smoking Gun Correlation II:\\
The Core Mass Correlation}
\label{knisec:smoke2}

One obvious explanation for the lack of convincing correlations
between global blue straggler numbers and cluster parameters is that
both the binary and the collisional channels contribute. In
particular, it seems plausible that each channel may dominate in
different regions within a cluster, with collisions perhaps dominating
in the dense core, and binary evolution dominating in the
periphery. More generally, it seems safe to assume that if
collisions/dynamics dominates blue straggler production anywhere, it
will be in cluster cores\index{cluster core}. So is it possible that a cleaner picture may
emerge if we focus specifically on blue stragglers found in the cores
of their parent clusters?

\begin{figure}
\begin{center} 
\includegraphics[width=119mm]{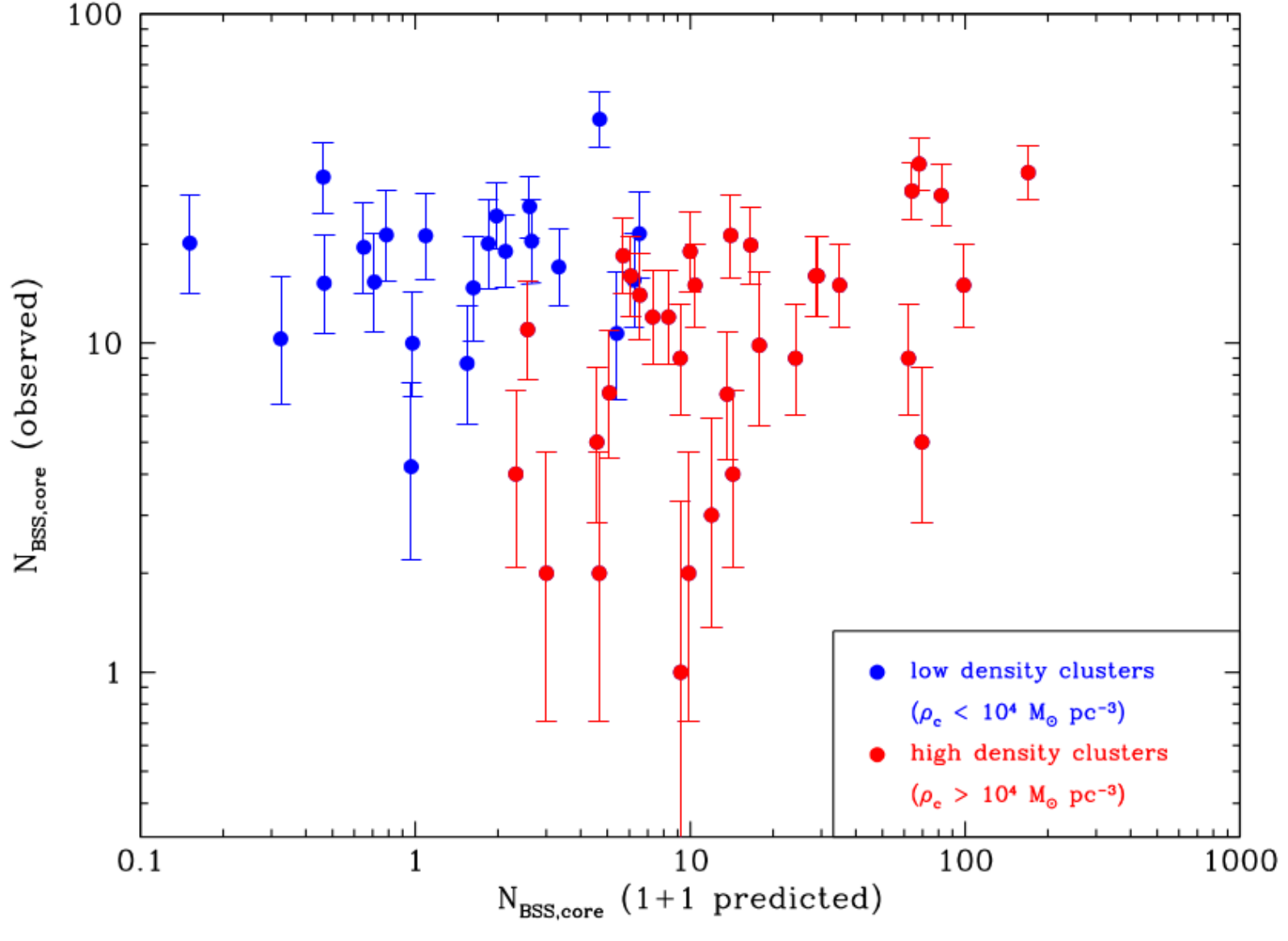}
   \caption{The number of blue stragglers found in a globular cluster
     core versus the number expected from 1+1 collisions. The latter
     is simply the product of the $\Gamma_{coll,1+1}$ and an assumed
     blue straggler lifetime (1~Gyr). Blue points correspond to 
     low-density clusters, red points to high-density clusters (see
     legend). Figure reproduced from Figure~1 of Knigge, Leigh \&
     Sills \cite{Kni09}, {\em A Binary Origin for Blue Stragglers in Globular
       Clusters}, Nature, 457, 288.}
   \label{kls1}
   \end{center}
\end{figure}

We investigated this idea in \cite{Kni09}, building 
on a new blue straggler catalogue constructed by Leigh, Sills \&
Knigge \cite{LSK07,LSK08}. This catalogue was still based on the WFPC2 data
set of Piotto et al. \cite{Pio02}, but included only systems found in
the cluster core by a consistent photometric selection
algorithm. Our hope and expectation was that the number of {\em core}
blue stragglers {\em would} show a strong correlation\index{correlation} with cluster
collision rate\index{collision rate}. However, Figure~\ref{kls1} shows that we were
wrong. More in desperation than expectation we then decided to also
have a look at the binary hypothesis. Since no comprehensive set of
empirically derived core binary fractions\index{core binary fraction}, $f_{bin,core}$, were
available in 2009, our only option was to use the total core mass,
$M_{core}$, as a proxy for the number of binaries in the core. This is
reasonable so long as core binary fractions do not vary dramatically
between clusters.

\begin{figure}
\begin{center} 
\includegraphics[width=119mm]{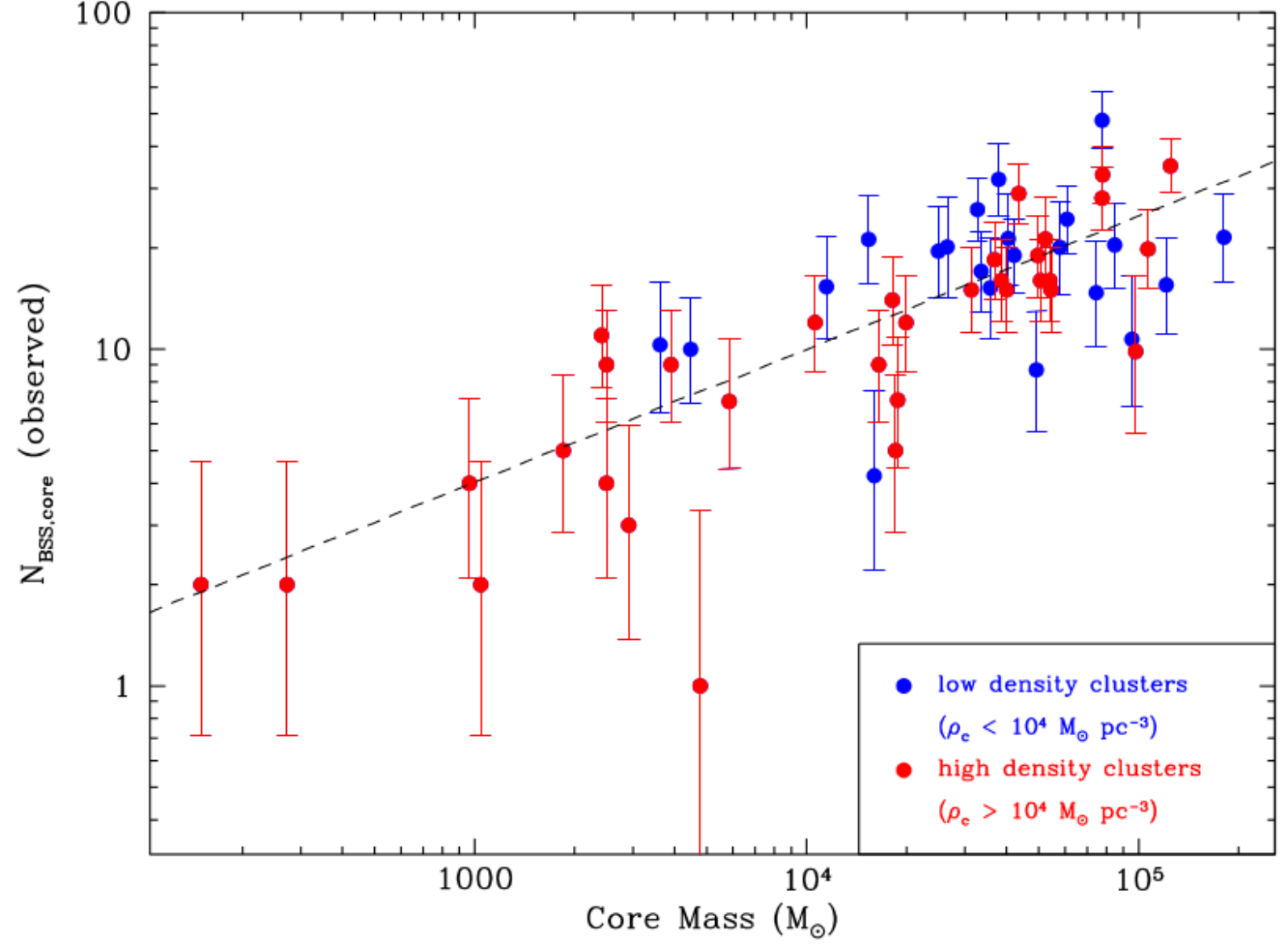}
   \caption{The number of blue stragglers found in a globular cluster
     core versus the total mass of the core. Blue points correspond to
     low-density clusters, red points to high-density clusters (see
     legend). Figure reproduced from Figure~2 of Knigge, Leigh \&
     Sills \cite{Kni09}, {\em A Binary Origin for Blue Stragglers in Globular
       Clusters}, Nature, 457, 288.}
   \label{kls2}
   \end{center}
\end{figure}

Much to our surprise, plotting $N_{BSS,core}$ vs $M_{core}$
immediately revealed a clear correlation (Figure~\ref{kls2}). This
would seem to suggest that stragglers are preferentially formed via
the binary channel\index{binary channel}, even in dense cluster cores. However, the observed
scaling is clearly sub-linear, and a fit to the data suggests
$N_{BSS,core} \propto M_{core}^{0.4}$. Can this be accommodated within
a simple binary scenario?

The simplest way to accomplish this is to remember that the intrinsic
scaling should be with the number of binaries in the core, not just
the core mass, i.e. $N_{BSS,core} \propto N_{bin,core} \propto
f_{bin,core} M_{core}$ in the binary picture. Thus the observed
scaling could be trivially understood if the core binary fractions
themselves scale with core mass as $f_{bin,core} \propto M_{core}^{-0.6}$. 

As already noted above, there was no definitive set of core binary
fractions available to test this prediction in 2009. However, Sollima
et al. \cite{SBFF07,SBFF08} had derived empirical binary fraction for a small sample
of low-density clusters and had already shown that these correlated
positively with the blue straggler frequencies in these clusters. Also,
Milone et al. \cite{Mil08} had just obtained preliminary estimates of the
core binary fractions in a larger sample of clusters, based on the
HST/ACS survey\index{HST/ACS survey} of Galactic globular clusters \cite{SBC07} and found a clear anti-correlation between core binary
fractions and total cluster mass. Even though neither set of core
binary fractions were suitable for combining directly with the blue
straggler data based used in \cite{Kni09}, they
permitted preliminary tests for a correlation with core mass. These
tests suggested that $f_{bin,core} \propto M_{core}^{-0.35}$, not too
far from the expected relation, albeit with considerable scatter.

In Section~\ref{knisec:smoke3}, we will consider whether more recent, higher
quality observations confirm, refute or modify these results. However, in
2009, our conclusion was that blue stragglers seem to be derived
mainly from binary systems, even in dense cluster cores\footnote{We
were careful not to rule out the possibility that the relevant binary
population may be affected by dynamical encounters\index{dynamical encounter}. However, we also
noted that the absence of a scaling with collision rates seemed hard to
understand in any scenario involving lots of dynamical encounters (see
Section~\ref{knisec:straw}).}.

\section{Alternative Constraints on Formation Channels}
\label{knisec:alternative}

Let us accept for the moment that the scaling of blue straggler
numbers with cluster parameters tends to favour a binary formation
channel. Are there other strands of evidence that would challenge this
idea? 

As discussed elsewhere in this book, it is
extremely difficult to confidently assign a specific formation
mechanism to a particular blue straggler. The only convincing cases
are the Carbon/Oxygen-depleted blue stragglers, which were initially
discovered by Ferraro et al. \cite{Fer06} in 47~Tuc\index{47 Tucanae}. This chemical anomaly\index{chemical anomaly}
is an expected consequence of mass transfer, since this process can
dredge up CNO-processed material from the stellar interior. By
contrast, no unique spectroscopic signature for
dynamically/collisionally formed blue stragglers is known. 

Nevertheless, there are at least two other types of observations that
may shed light on blue straggler formation in globular clusters. They
are (i) the radial distribution\index{radial distribution} of blue stragglers in a given cluster,
and (ii) the discovery of a double blue straggler sequence in M30\index{M30} (and
perhaps other clusters). Both of these observations are discussed in
much more detail elsewhere in this book, so here we will merely ask 
whether (or to what extent) they conflict with the idea that most blue
stragglers derive from binaries, rather than from dynamical encounters. 

\subsection{Radial Distributions}
\label{knisec:radial}

\begin{figure}
\begin{center} 
\includegraphics[width=119mm]{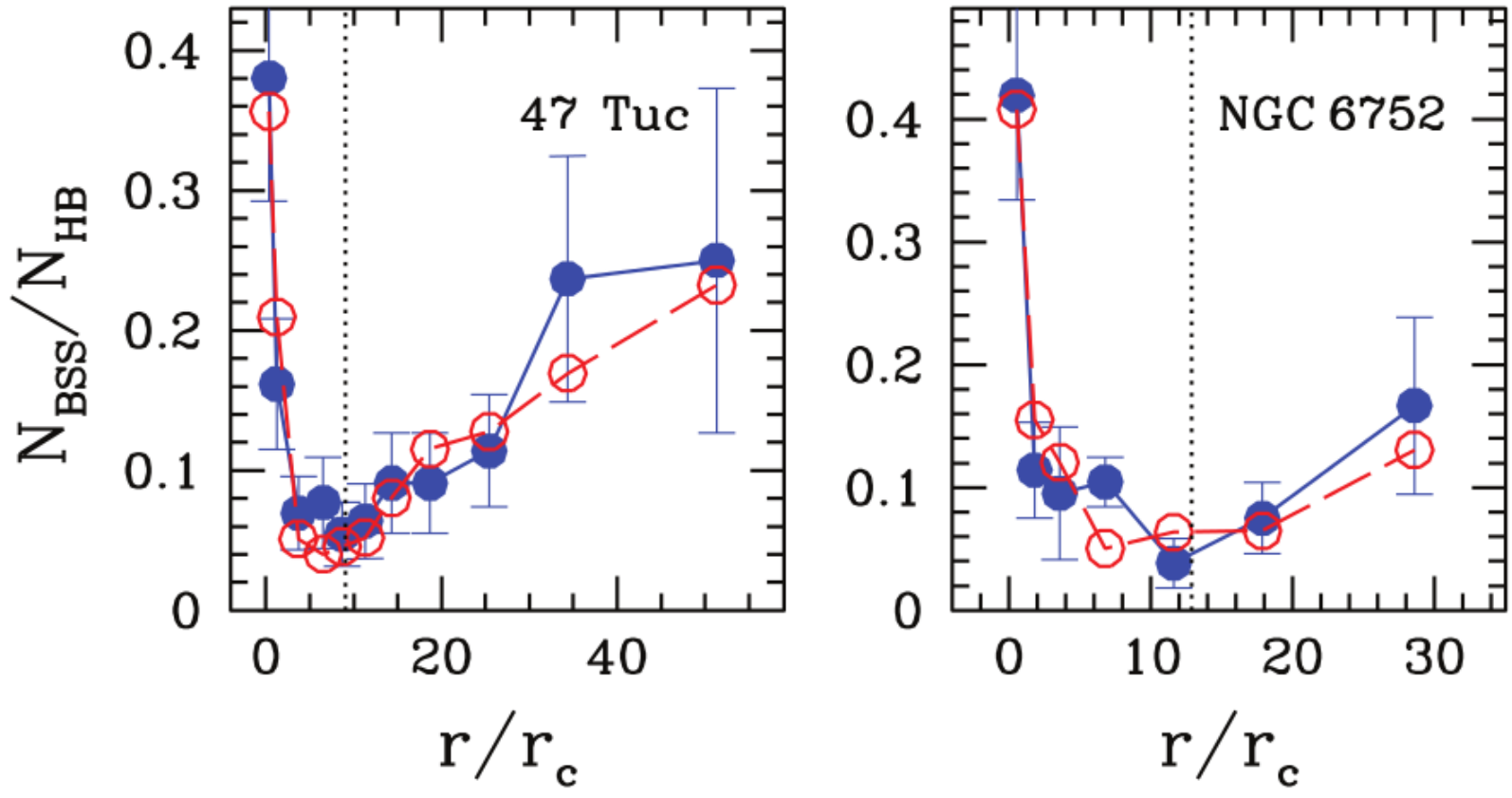}
   \caption{The radial distribution of blue straggler frequencies in
     the globular clusters 47~Tuc and NGC~6752. Blue straggler frequencies are
     defined here as the number of blue stragglers normalised to the
     number of horizontal branch stars. The filled blue point
     and solid blue lines are the observational data, while the open
     red circles and red dashed lines are the best-fitting models of
     Mapelli et al. \cite{Ma06}. The dotted lines mark the location of the
     minimum in the radial distributions. See text for details. Figure
     adapted from Figure 2 of Mapelli et al. \cite{Ma06}, {\em The Radial
       Distribution of Blue Straggler Stars and the Nature of their
       Progenitors}, MNRAS, 373, 361.}
   \label{mapelli}
   \end{center}
\end{figure}

In most globular clusters, the dependence of blue straggler {\em
frequency} on radius is bimodal (Figure~\ref{mapelli}; e.g. \cite{Fer97,SFS04,WSB06,Lan07a,Lan07b}. These distributions have been modelled quite
successfully by Mapelli et al. \cite{Ma04,Ma06}. Their simulations follow
the motion of blue stragglers in a static cluster potential, assuming
that collisional blue stragglers form only within the core, while
binary-derived blue stragglers all start their lives outside the
core. The ratio of collisional to binary blue stragglers is a free
parameter of the model.

Mapelli et al. \cite{Ma04,Ma06} obtained the best fits to the observed
distributions with both channels contributing a comparable number of
blue stragglers to the total population. Moreover, they also found
that collisional blue stragglers completely dominate the population
in the cluster core\index{cluster core}, while binary blue stragglers are dominant in the
cluster halo\index{cluster halo}, beyond the minimum in the radial distribution. Yet this
conclusion seems incompatible with the lack of a correlation between
core blue straggler numbers and cluster collision rates\index{collision rate}.

It is useful to take a step back at this point and consider the
physics that produces the bimodal blue straggler
distributions. As discussed by Mapelli et al. \cite{Ma04,Ma06} and
described in more detail elsewhere in this book, the key
dynamical process is dynamical friction. Since blue stragglers are
relatively massive, they tend to sink towards the cluster core. The
time scale on which this happens, $t_{df}$, changes as a function of
radius. We can therefore define a critical radius, $R_{min}
\simeq R(t_{dc} = t_{gc})$, where $t_{gc}$ is the lifetime of the
cluster. Binary blue stragglers born well inside $R_{min}$ have had
plenty of time to sink to the core, while those born well outside this
radius have barely moved from their original location. The minimum in
the blue straggler distribution therefore corresponds roughly to
$R_{min}$.

These considerations highlight an important point: a blue straggler
population containing {\em only} binary blue stragglers should still
produce a bimodal radial distribution. So why
are collisional blue stragglers needed at all in the simulations?
The answer is that not enough binary blue stragglers were seeded
inside $R_{min}$. But this is just an assumption. If the birth
distribution of binary blue stragglers is allowed to be centrally
peaked, a population consisting exclusively of such systems may be
able to match the data as well (see \cite{Ma06}).

It is perhaps also worth emphasising here that the scenario
preferred by Mapelli et al. \cite{Ma04,Ma06} is very different from that
suggested by Davies et al. \cite{DPd04} . Both scenarios do favour a
combination of binary-derived and collisional blue
stragglers. However, in Davies et al. model, different
formation channels dominate in different {\em clusters}, whereas in
Mapelli et al.'s model, different channels operate in
different locations {\em within} a given cluster. In any case, the
key point for our purposes here is that a bimodal radial
distribution does not necessarily require distinct formation channels
for the core and halo blue straggler populations. 

\subsection{Double Blue Straggler Sequences}
\label{knisec:double_seq}

One other recent discovery is highly relevant to the question of
blue straggler formation channels. Ferraro et al. \cite{Fe09} showed that 
colour magnitude diagram of the globular cluster M30 appears to
contain {\em two} distinct blue straggler sequences
(Figure~ 5.7 in Chapter 5). Their interpretation of this observation is that 
objects on the blue sequence were formed via collisions, while those
on the red sequence are derived from binaries. 


Why should there be such a clean separation between these two types of
systems in M30? The idea put forward by Ferraro et al. \cite{Fe09} is that,
in M30, the collisional blue stragglers all formed recently in a short
burst, most likely when the cluster underwent core collapse. All of
these objects therefore share the same evolutionary state, so that all
of them line up on a well-defined main sequence. By contrast, the red
sequence lies roughly 0.75 mag above the extension of the cluster
zero-age main sequence, as expected for a population of roughly
equal-mass binaries. If this idea is correct, then both sequences
could be present in many/most clusters, but would usually overlap too
much to be noticeable as distinct entities.

The double blue straggler sequence in M30 is almost certainly an
important clue, and Ferraro et al. present new data elsewhere in this
book (see Chap. 5) that appear to show a similar double 
sequence in another cluster. If confirmed, it would be nice if each of
the two sequences really does correspond to a distinct formation
channel, even if this may make it harder to understand other results,
such as the core mass correlation. However, there is at least one
surprising aspect to the double sequence in M30. As noted by Ferraro
et al. \cite{Fe09}, their blue straggler sample for this cluster contains 3
W UMa binaries\index{W UMa star} and two other variables that are likely
binaries. However, these are not all located on the red (binary)
sequence. Rather, one W UMa and one other binary are located nicely on
the collisional sequence. Perhaps this simply means that these two
binaries were produced in (or affected by) dynamical encounters\index{dynamical encounter}, while
the others are mostly primordial\index{primordial binary}. Nevertheless, if each sequence
corresponds cleanly to a particular formation channel, it does seem
surprising that the known binaries should be split nearly evenly
between them.

\begin{figure}
\sidecaption
\includegraphics[width=69mm]{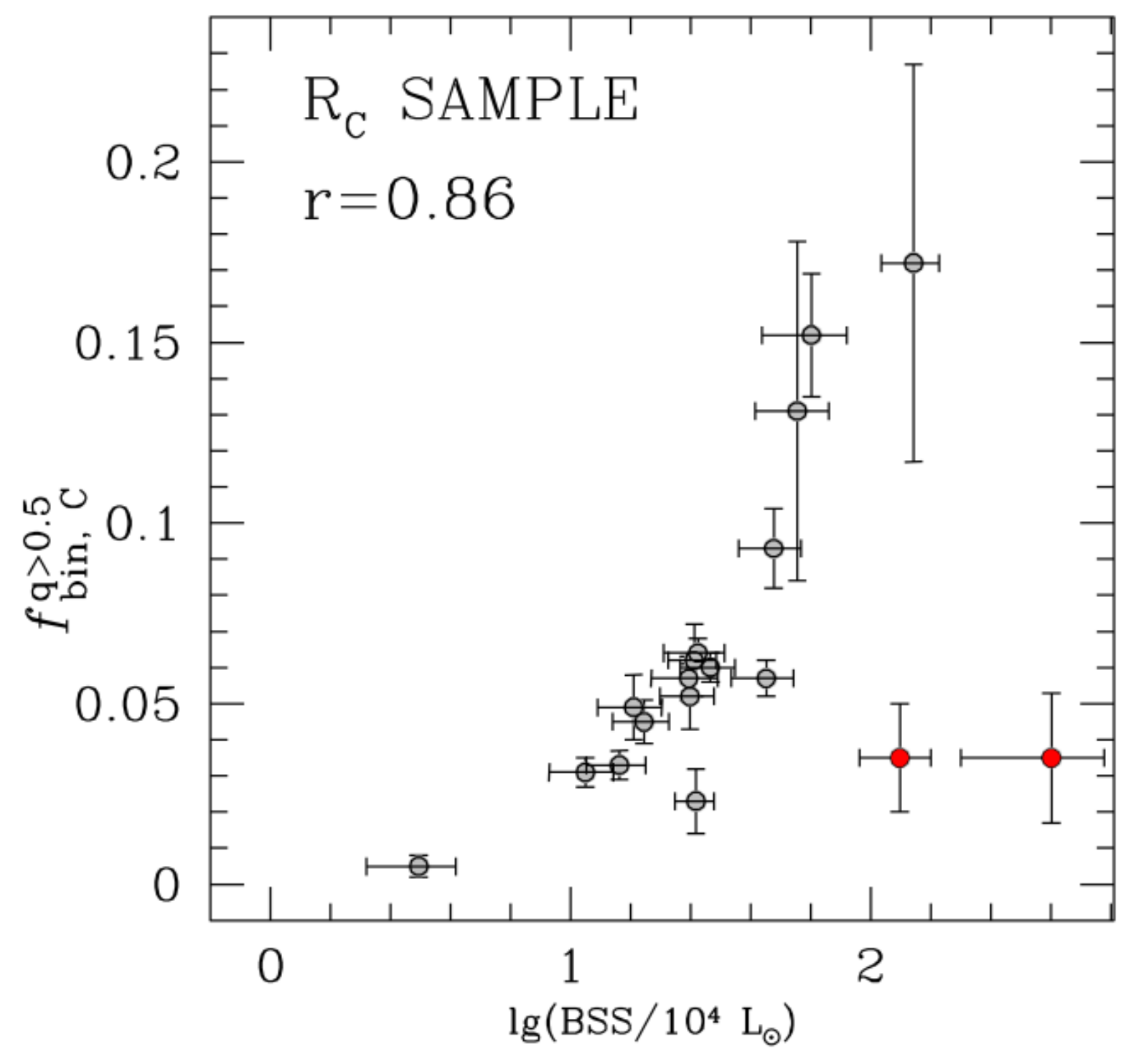}
   \caption{The fraction of binaries with mass ratios $q > 0.5$ in the
     cluster core versus the blue straggler frequency in the core. The
     two red points near the bottom right of the plot correspond to 
     post-core-collapse clusters. Figure reproduced from Figure~B5 of Milone et
     al. \cite{Mil12}, {\em The ACS Survey of Galactic Globular Clusters XII:
       Photometric Binaries along the Main Sequence}, A\&A, 540, A16.}
   \label{milone}
\end{figure}

\section{The Search for the Smoking Gun Correlation III:\\
Once More, With Binary Fractions...}
\label{knisec:smoke3}

Binaries are key to the study of cluster dynamics. In fact, the late
dynamical evolution of globular clusters is thought to be {\em driven}
by binary systems (e.g. \cite{Hut92}). It was therefore a major
breakthrough when Milone et al. \cite{Mil12} presented photometric estimates
of binary fractions for 59 clusters, based on the HST/ACS survey\index{HST/ACS survey} of
Galactic globular clusters already mentioned in Section~\ref{knisec:smoke2}.

Three trends discovered by Milone et al. are of immediate
relevance to the blue straggler formation problem. First, core binary
fractions correlate only weakly with $\Gamma_{coll,1+1}$. Second, they
anti-correlate more strongly with total cluster luminosity (and hence
mass). Third, they correlate very strongly with blue straggler {\em
frequency} (Figure~\ref{milone}). All of these trends are quite promising 
for the idea that binaries dominate blue straggler production, as
suggested in \cite{Kni09}.

However, the availability of binary fractions makes it possible to
test the key formation scenarios much more directly. For example, we
can now compare the number of core blue stragglers directly to 
the number of binaries in the core, rather than just to the total core
mass. Similarly, we can now directly estimate 1+2 and 2+2 encounter rates
($\Gamma_{1+2}$ and $\Gamma_{2+2}$; see
Section~\ref{knisec:straw}). If blue straggler production is dominated by
encounters involving binaries 
(e.g. exchange encounters; see Section~\ref{knisec:burn}), blue straggler
numbers should correlate strongly with $\Gamma_{1+2}$ or
$\Gamma_{2+2}$.

We carried out these tests in \cite{Leigh13}. For this purpose,
we combined the core blue straggler numbers derived by Leigh et
al. \cite{LSK11} with the core binary fractions obtained by Milone et
al. \cite{Mil12}. These data sets are ideally matched, since both are based
on the HST/ACS survey of Galactic globular clusters \cite{SBC07}.

\begin{figure}
\begin{center} 
\includegraphics[width=49mm]{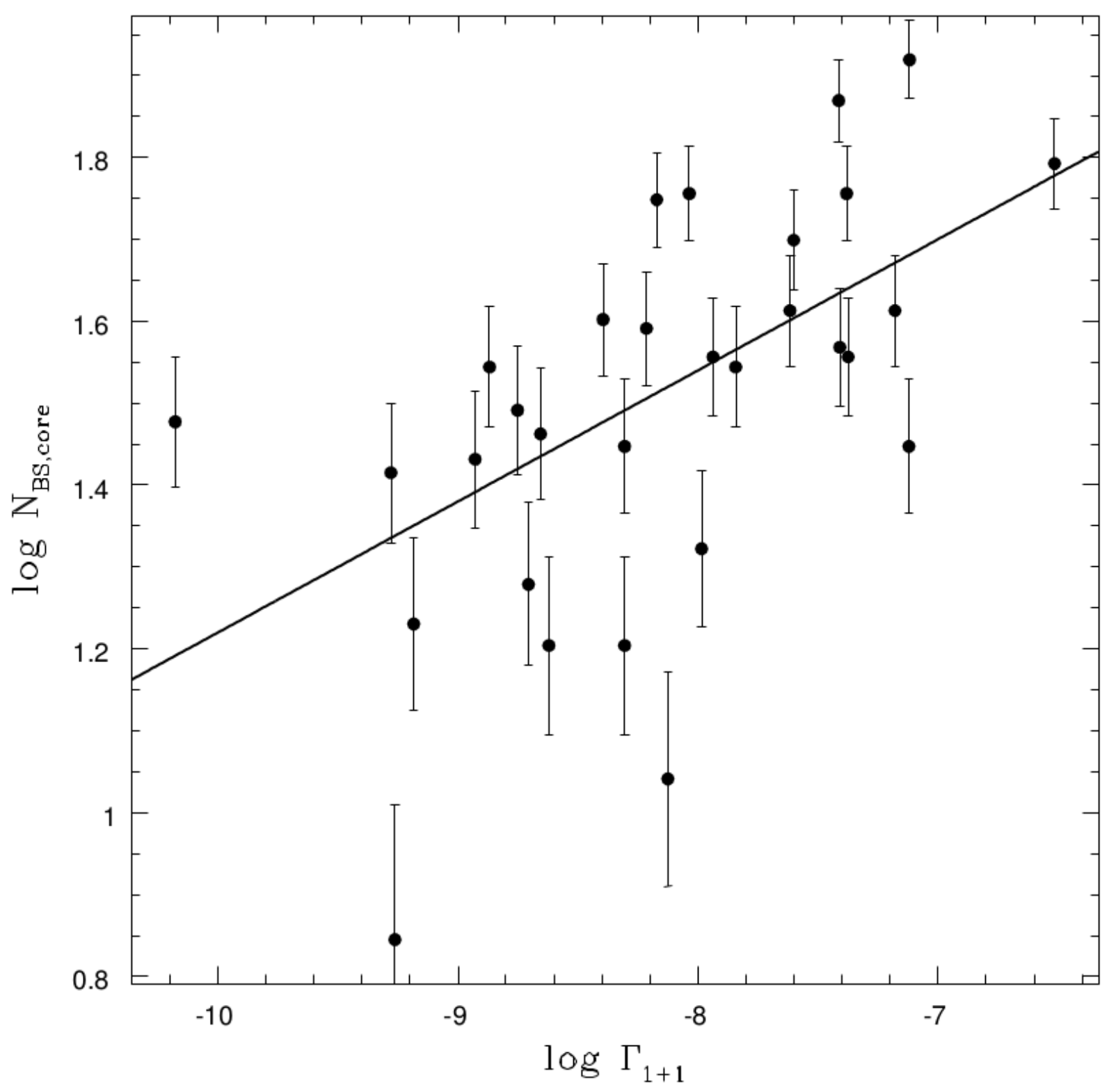}
\includegraphics[width=49mm]{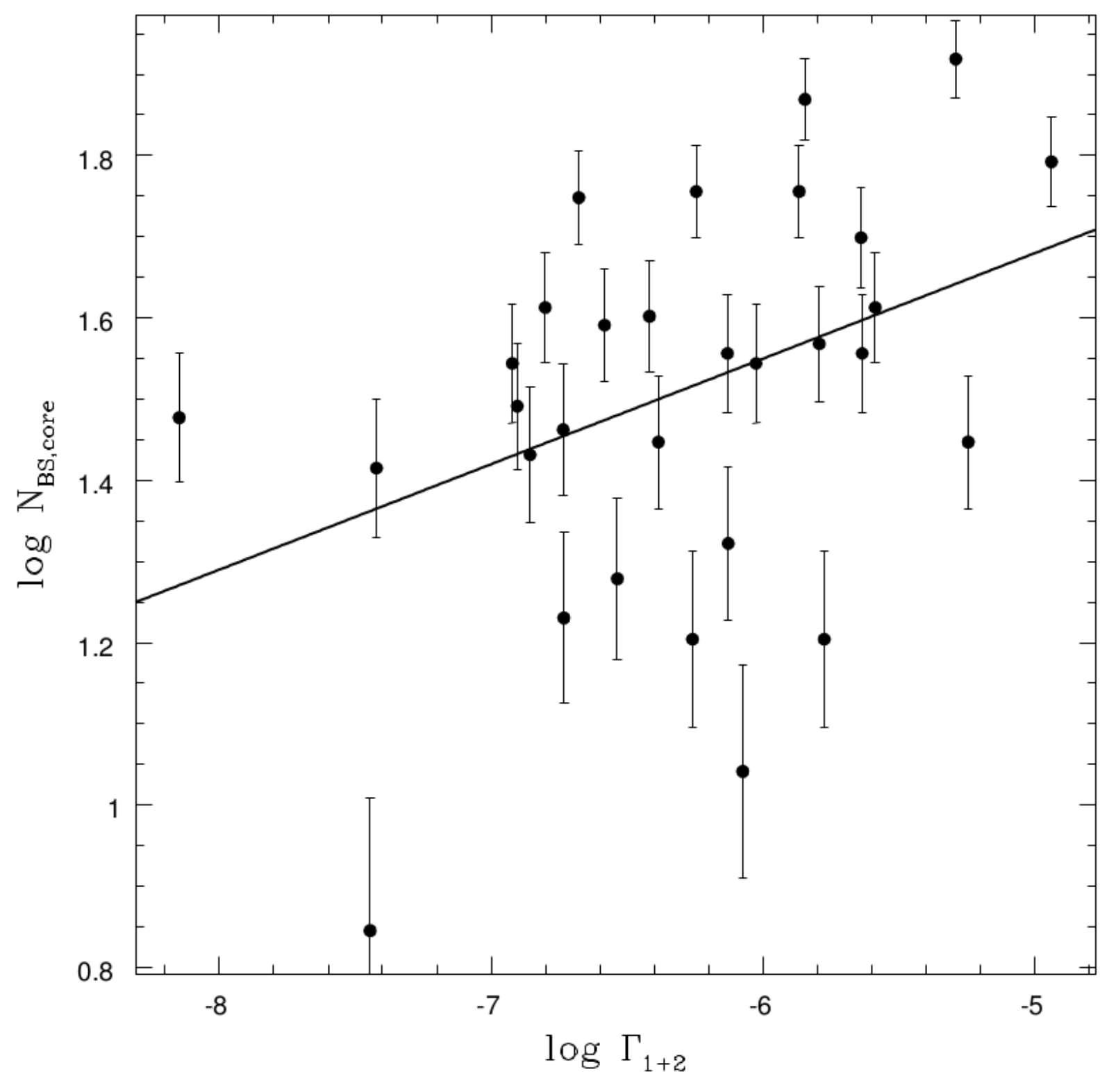}
\includegraphics[width=49mm]{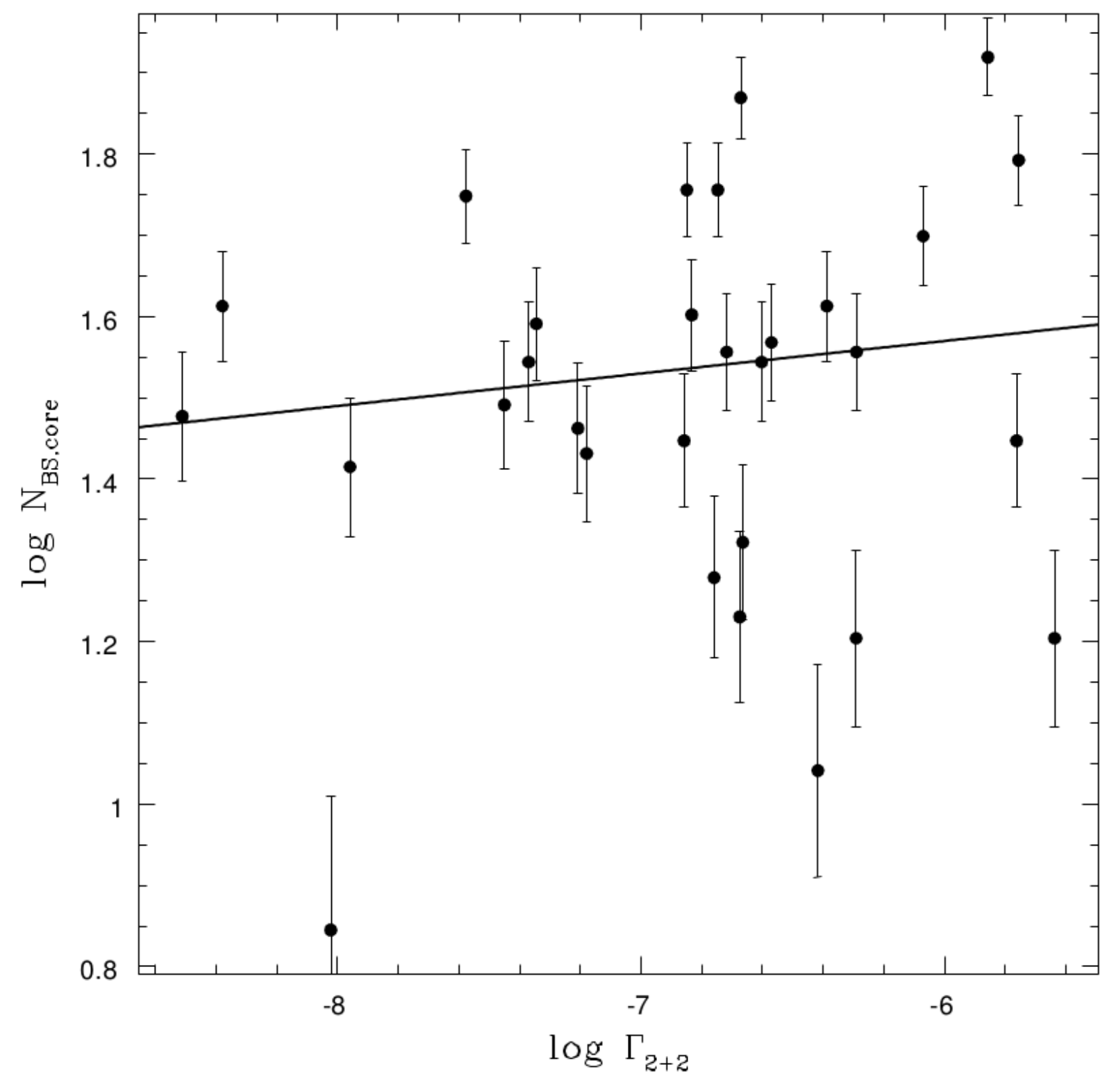}
   \caption{The number of blue stragglers in the core versus the 1+1
     collision rate (top left panel), the 1+2 encounter rate (top
     right panel)
     and the 2+2 encounter rate (bottom panel). Figure reproduced/adapted from
     Figures 6, 7 and 8 of Leigh et al. \cite{Leigh13}; {\em The Origins of Blue
       Straggler and Binarity in Globular Clusters}, MNRAS, 428, 897.}
   \label{leigh1}
   \end{center}
\end{figure}

Let us first look at the results for dynamical formation
scenarios. Figure~\ref{leigh1} shows plots of $N_{BSS,core}$ against each
of $\Gamma_{coll,1+1}$ (top left panel), $\Gamma_{1+2}$ (top
right panel) and $\Gamma_{2+2}$ (bottom panel). None of the encounter
rates correlate cleanly with the number of blue stragglers in cluster
cores\footnote{As already noted by Leigh et al. \cite{LSK11}, the correlation
between $N_{BSS,core}$ and $\Gamma_{coll,1+1}$ is formally
significant, but probably induced by  the intrinsic correlation
between $\Gamma_{coll,1+1}$ and $M_{core}$ (see Section~\ref{knisec:straw}).}.

\begin{figure}
\begin{center} 
\includegraphics[width=49mm]{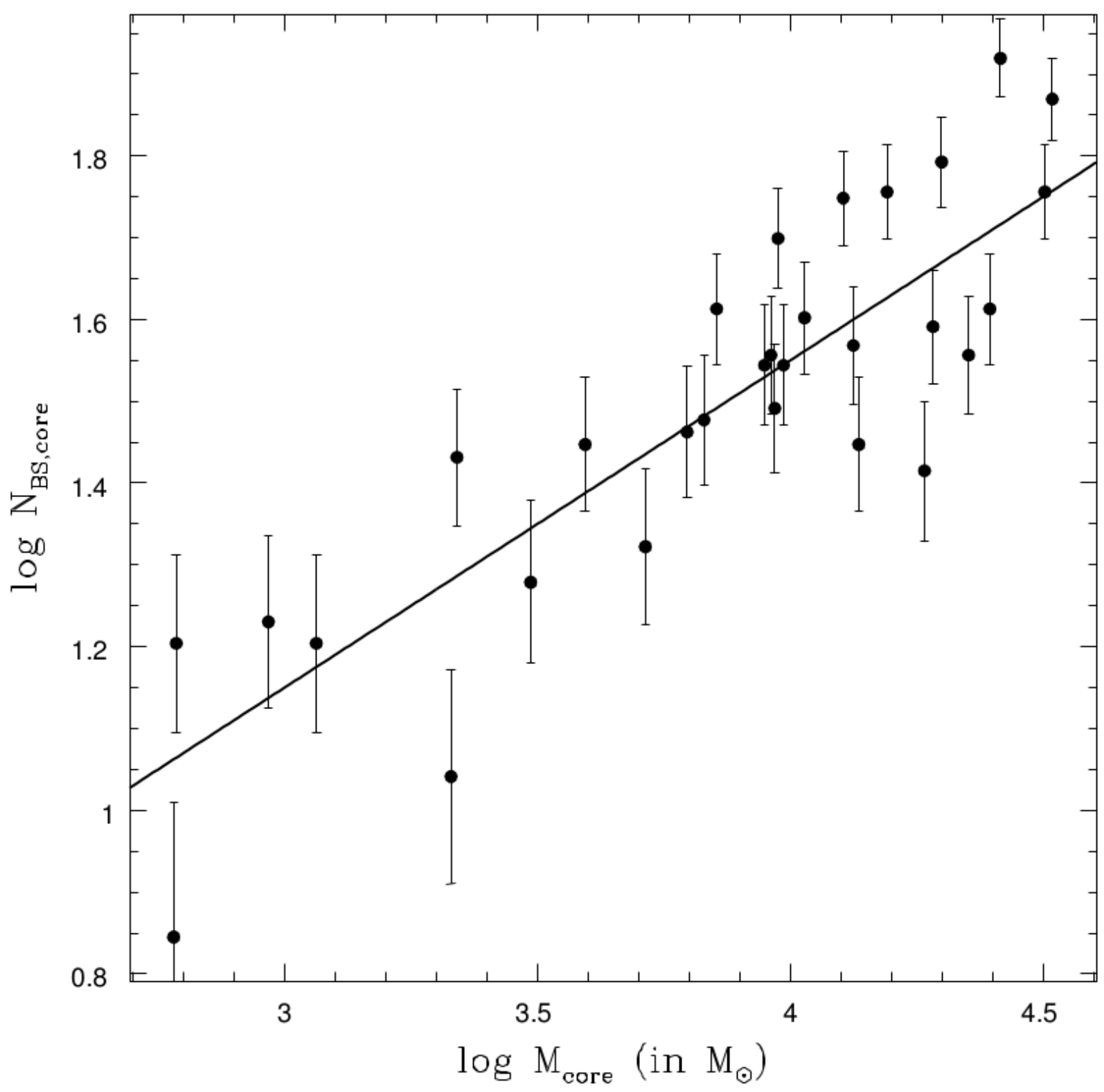}
\includegraphics[width=49mm]{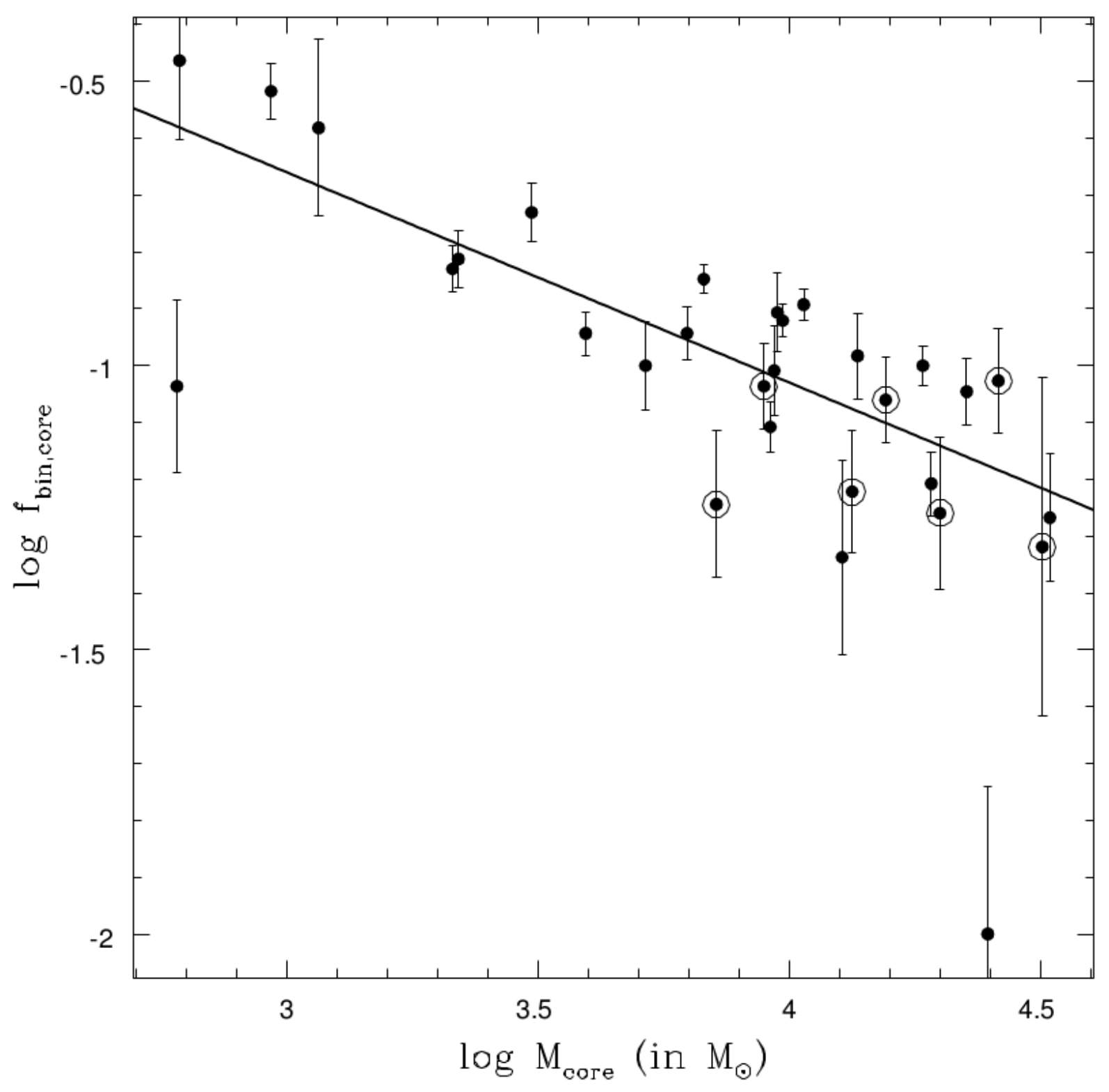}
\includegraphics[width=49mm]{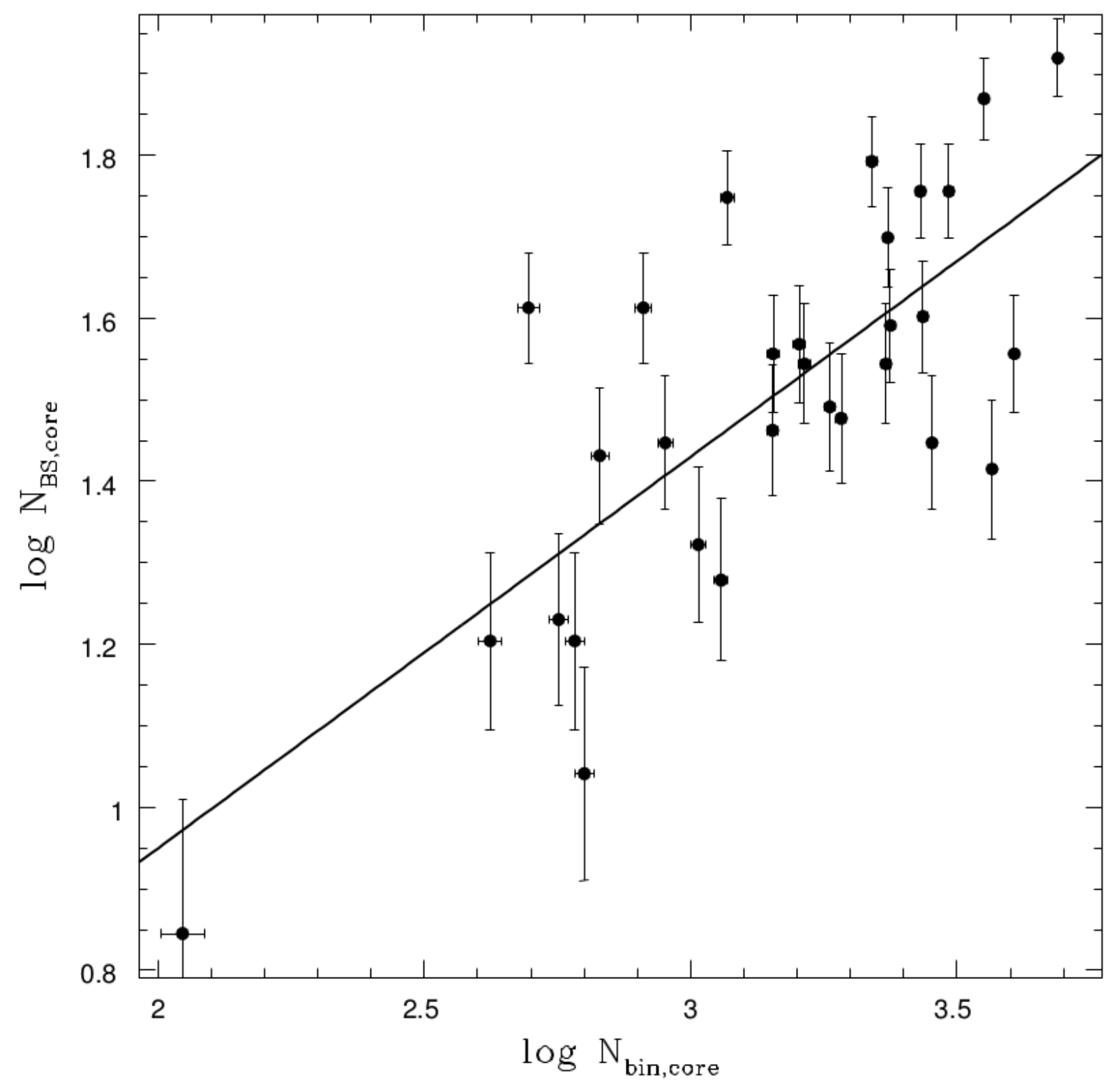}
 \caption{Diagnostic plots for the binary evolution scenario for
     blue straggler formation, based on the HST/ACS survey of
     Galactic globular clusters (Sarajedini et al. 2007).  
     {\em Top left panel:} The number of blue stragglers in the core
     versus the mass of the core. {\em Top right panel:} Core binary
     fraction versus core mass. {\em Bottom panel:} The number of blue
     stragglers in the core versus the estimated number of binaries in
     the core. Figure reproduced/adapted from
     Figures 2, 3 and 4 of Leigh et al. \cite{Leigh13}; {\em The Origins of Blue
     Straggler and Binarity in Globular Clusters}, MNRAS, 428, 897.}
   \label{leigh2}
   \end{center}
\end{figure}

Now let us look at the binary evolution scenario. The top left panel in
Figure~\ref{leigh2} shows that the ACS data confirms the existence of a
strong correlation between $N_{BSS,core}$ and core mass, with $N_{BSS}
\propto M_{core}^{0.4}$ (also see \cite{LSK11}). The top right panel
in Figure~\ref{leigh2} shows that the data also confirm the prediction of
a strong {\em anti-correlation} between core binary fraction and
$M_{core}$. A power-law fit to this relation
gives roughly $f_{bin,core} \propto M_{core}^{-0.4}$, a little
shallower than predicted, but not too far from the expected
$M_{core}^{-0.6}$ dependence. So far, so promising. However, the
bottom panel in Figure~\ref{leigh2} shows what happens when we directly 
compare $N_{BSS,core}$ to $N_{bin,core} \propto f_{bin,core}
M_{core}$. Instead of improving on the correlation with core mass
alone, the addition of empirical binary fractions actually {\em
degrades} it!

This is a surprising result. One possibility is that it simply means
that all of our straw-man models are too simplistic after all
(although, as noted in Section~\ref{knisec:straw}, it seems quite hard to avoid
all of the expected correlations, even in more complex formation
scenarios). However, before we accept that we need ``new physics'', we 
should check if we can somehow reconcile one of the existing models
with our new findings. 

Since the binary evolution model predicts at least the observed
correlation with core mass\index{core mass} (and the lack of a correlation with
collision rate\index{collision rate}), let us imagine that all blue stragglers are
exclusively formed from (primordial) binaries\index{primordial binary}. In this case,
$N_{BSS,core} \propto N_{bin,core} \propto f_{bin,core} M_{core}$,
with just some modest intrinsic scatter. We already know
empirically that $f_{bin,core}$ anti-correlates quite strongly with
$M_{core}$. But now suppose that the intrinsic anti-correlation is
even stronger than the observed one, i.e. that the scatter in the
middle panel of Figure~\ref{leigh2} is mostly due to observational
errors on $f_{bin,core}$, rather than any intrinsic dispersion. In this limit, {\em
$M_{core}$ actually becomes a better predictor of the true core binary
fractions than the observationally estimated values.} The number of
binaries in the core -- and hence the number of blue stragglers --
will then also be predicted more accurately by $M_{core}$ alone than
by the empirically estimated combination of $f_{bin,core}M_{core}$. 

We have carried out some simple simulations to test and illustrate
this idea. In these simulations, we create mock data sets\index{mock data set} of similar
size and dynamic range as the real data and assume that the number of
blue stragglers scales perfectly and linearly with the number of
binaries, i.e. $N_{BSS,core} \propto N_{bin,core}$. We also assume
that $f_{bin,core} \propto M_{core}^{0.6}$, with only a slight
intrinsic dispersion, $\sigma_{int}$. Finally, we assume that our
observational estimates of $f_{bin,core}$ are subject to an
observational uncertainty of $\sigma_{obs}$, which we vary in the
range $0.1\sigma_{int} \leq \sigma_{obs} \leq 10.0 \sigma_{int}$. We
then analyse each mock data set to estimate the correlation
coefficients of $N_{BSS,core}$ against the ``observationally
estimated'' $M_{core}$ and $N_{bin,core} = f_{bin,core} M_{core}$. We
also fit the latter correlation with a power law and estimate the
power law index.

\begin{figure}
\begin{center} 
\includegraphics[width=119mm]{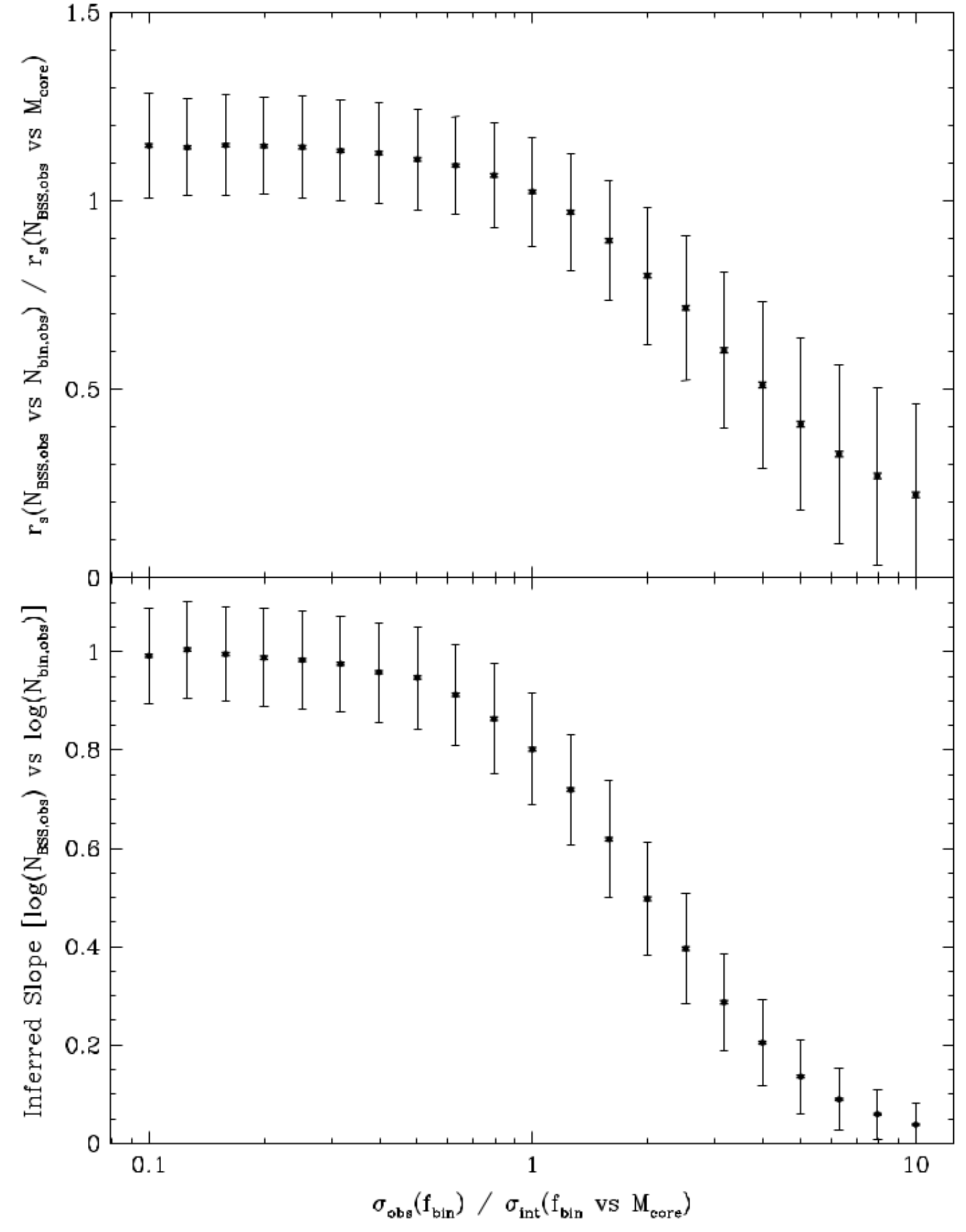}
   \caption{The effect of observational uncertainties in binary
     fractions on correlations between blue straggler 
     numbers and cluster parameters. The data shown in both panels are
     derived from simulations designed to roughly mimic the data shown
     in Figure~\protect{\ref{leigh2}} (see text for details). The
     ordinate in both panels is the ratio of the assumed observational
     uncertainty, $\sigma_{obs}$, to the assumed intrinsic dispersion
     in the relation between $f_{bin,core}$ and $M_{core}$. 
     {\em Top panel:} The ratio of the
     correlation coefficients between $N_{BSS,core}$ and
     $N_{bin,core}$, on the one hand, and $N_{BSS,core}$ and
     $M_{core}$, on the other; note that the correlation with
     $N_{bin,core}$ will only seem stronger than that with $M_{core}$
     if $\sigma_{obs}<<\sigma_{int}$. 
     {\em Bottom panel:} The inferred power 
     law index of the $N_{BSS,core}$ versus $N_{bin,core}$
     correlation; note that the correct value of unity is only
     obtained if $\sigma_{obs}<<\sigma_{int}$. Figure reproduced from
     Figures 9 of Leigh et al. \cite{Leigh13}; {\em The Origins of Blue
       Straggler and Binarity in Globular Clusters}, MNRAS, 428, 897.}
   \label{leigh3}
   \end{center}
\end{figure}

Figure~\ref{leigh3} shows how the ratio of the estimated correlation
coefficients, and also the inferred power law index of the
$N_{BSS,core}$ vs $N_{bin,core}$ relation, depend on the ratio of
$\sigma_{obs}/\sigma_{int}$. As expected, when $\sigma_{obs} <<
\sigma_{int}$, the
``empirical'' core binary fractions add value. In this case, the
correlation\index{correlation} coefficient between $N_{BSS,core}$ and $N_{bin,core}$ is
larger than that between $N_{BSS,core}$ and $M_{core}$. Also, the
inferred power law index of the $N_{BSS,core}$ vs $N_{bin,core}$
relation is unity, i.e. we correctly infer that the intrinsic relation
is linear. However, when $\sigma_{obs} >> \sigma_{int}$, the
empirical binary fractions only serve to degrade the underlying
signal. In this limit, the correlation of $N_{BSS}$ with $M_{core}$ is 
stronger than that with the estimated $N_{bin,core}$, and the
relationship between $N_{BSS,core}$ and $N_{bin,core}$ is incorrectly
inferred to be sub-linear. So the binary evolution {\em might} still
be consistent with the observations, {\em but only if (core) binary
fractions correlate extremely cleanly with cluster (core) masses.}

\section{Summary \& Outlook}
\label{knisec:summ}

What are the key points to take away from our look at blue straggler
statistics? On the observational front, we have seen that (i) 
blue straggler numbers do {\em not} correlate with dynamical encounter
rates; (ii) they {\em do} correlate strongly with cluster (core)
masses; (iii) empirically estimated core binary numbers (obtained by
combining core masses with photometrically determined core binary
fractions) correlate {\em less} strongly with blue straggler numbers
than core masses alone.

The first point would seem to argue against a dynamical origin for
most blue stragglers in globular clusters. Yet presumably encounters
and collisions must happen in such dense environments at roughly the
predicted rates. So can the efficient production of blue stragglers
via this channel actually be avoided? Sills et al. \cite{Si01} followed the
 evolution of a simulated stellar collision products. One of their
key findings was that, if left to their own devices, such objects tend
to exceed the break-up velocity\index{break-up velocity} and can be completely disrupted if
unbound mass shells are successively removed from the surface. In
their words: ``either blue stragglers are not created through physical
off-axis collisions or some mechanism(s) can remove angular momentum
from the star on short timescales'' \cite{Si01}. Thus perhaps
stellar collisions\index{stellar collision} occur, but do not produce blue stragglers. 

On the other hand, recent dynamical simulations of globular clusters
suggest that, even if blue stragglers are produced predominantly by
(mainly binary-mediated) collisions, their numbers may scale only
weakly with $\Gamma_{coll,1+1}$ \cite{Chat13,Si13}. The reason for this is not immediately apparent, however,
and it is also not clear if a strong scaling with $\Gamma_{1+2}$
could be avoided as well. 

The second point -- the core mass correlation -- can be explained
most naturally in the context of a binary scenario for blue
straggler formation\footnote{We should note, however, that Chatterjee et
al. \cite{Chat13} and Sills et al. \cite{Si13} argue that the dynamically-formed
blue stragglers in their simulations also produce a core mass
correlation.}.
However, the third point -- the poor correlation
obtained when core masses are combined with empirically estimated
binary fractions -- seems at first sight inconsistent with a binary 
scenario. We have seen that this discrepancy can be 
resolved if core binary fractions\index{core binary fraction} are extremely tightly coupled to
core masses. If this idea is correct, it would have significant
implications for our understanding of cluster dynamics, well 
beyond the realm of blue stragglers. 
 
\begin{acknowledgement}
I am extremely grateful to the organisers of the ESO workshop that led
to the production of this book. Special thanks are due to Henri
Boffin, whose patience as an editor was almost literally unlimited.
\end{acknowledgement}

\backmatter
\printindex


\end{document}